\documentclass{article}
\usepackage{arxiv}

\author{
        Andrew Mellor \\
        Mathematical Institute\\
		University of Oxford\\
		\texttt{mellor@maths.ox.ac.uk} 
}
\date{\today}

\usepackage{amsmath, amssymb}
\usepackage{graphicx}
\usepackage{xcolor}
\usepackage{hyperref}
\usepackage{bm}
\usepackage{subcaption}
\usepackage{booktabs}

\renewcommand{\vec}[1]{\bm{#1}}

\usepackage{amsthm}

\theoremstyle{definition}
\newtheorem{definition}{Definition}[section]

\title{Event Graphs: Advances and Applications of Second-Order Time-Unfolded Temporal Network Models}

\begin{document}

\maketitle

\begin{abstract}
	\noindent Recent advances in data collection and storage have allowed both researchers and industry alike to collect data in real time.
	Much of this data comes in the form of `events', or timestamped interactions, such as email and social media posts, website clickstreams, or protein-protein interactions.
	This of type data poses new challenges for modelling, especially if we wish to preserve all temporal features and structure.
	We propose a generalised framework to explore temporal networks using second-order time-unfolded models, called \emph{event graphs}.
	Through examples we demonstrate how event graphs can be used to understand the higher-order topological-temporal structure of temporal networks and capture properties of the network that are unobserved when considering either a static (or time-aggregated) model.
	Furthermore, we show that by modelling a temporal network as an event graph our analysis extends easily to consider non-dyadic interactions, known as hyper-events.	

	\vspace{1em} \noindent \textbf{Keywords:} temporal networks, higher-order models, hypergraphs, percolation 
\end{abstract}

\section{Introduction}
\label{sec:introduction}

As an abstraction of complex systems, networks have been a fundamental tool for research across a number of disciplines.
Typically network models assume that the relationships between objects (nodes) are static or unchanging.
The static network assumption is usually made for either ease of analysis, or simply because of the lack of temporal data.
This is particularly the case in the biological sciences where data collection can be both difficult and costly.

In recent years our capacity to collect data has improved drastically, especially in the digital domain where many systems can be monitored autonomously.
This means that for many applications individual timestamped interactions between objects (also called \emph{events}) are recorded.
In social networks events may take for the form of messages between users, or in biology an event may be the interaction between two proteins at a certain time.
In some cases the events are not timestamped but are temporally ordered. 
This type of data falls under the wider category of sequential data.
The modelling and analysis of these timestamped events (or sequential data) has become a vital task with applications in data mining, social network analysis, and computational biology.

Despite static networks providing a useful abstraction of complex systems, many of these systems themselves are not stationary and the interactions between objects change over time or appear only in a particular order~\cite{holme2013temporal}.
Systems that evolve over time can instead be modelled as a \emph{temporal network}.
While most systems exhibit at least some temporal dependency there are relatively few tools to capture their underlying topological-temporal behaviour~\cite{masuda2016guide,holme2013temporal,holme2015modern}.
There are many ways to represent temporal networks however we focus on the event-based representation~\cite{kempe2000connectivity}.
In this representation a temporal network is described by a time-ordered sequence of temporal events $(e_i)_{i=1}^M$ where each event is of the form $e_i = (u_i,v_i,t_i,\delta_i)$.
Here $u_i$ and $v_i$ are the interacting node pair of the $i$th event which occurs at time $t_i$ and lasts for a duration $\delta_i$.
If the network is directed this is interpreted as a directed event from $u_i$ to $v_i$.
Furthermore the duration of events is often omitted or is set to zero for instantaneous interaction.

Studies of temporal networks typically revolve around the timings between events occurring, or the \emph{inter-event time}~\cite{navaroli2015modeling}.
How these times are distributed have been shown to have an effect on dynamics evolving across the system~\cite{scholtes2014causality,masuda2013predicting,masuda2013temporal}.
Other studies have investigated the prevalence of temporal motifs in these systems~\cite{kovanen2011temporal,kovanen2013temporal,paranjape2017motifs}.
Network motifs are small repeated patterns of interaction in the network and their application in the temporal setting has perhaps shown the most promise in characterising behaviour in temporal networks.

\subsubsection*{Time-unfolded Graphical Models}

\begin{figure}[!h]
	\centering
	\includegraphics[width=0.7\linewidth]{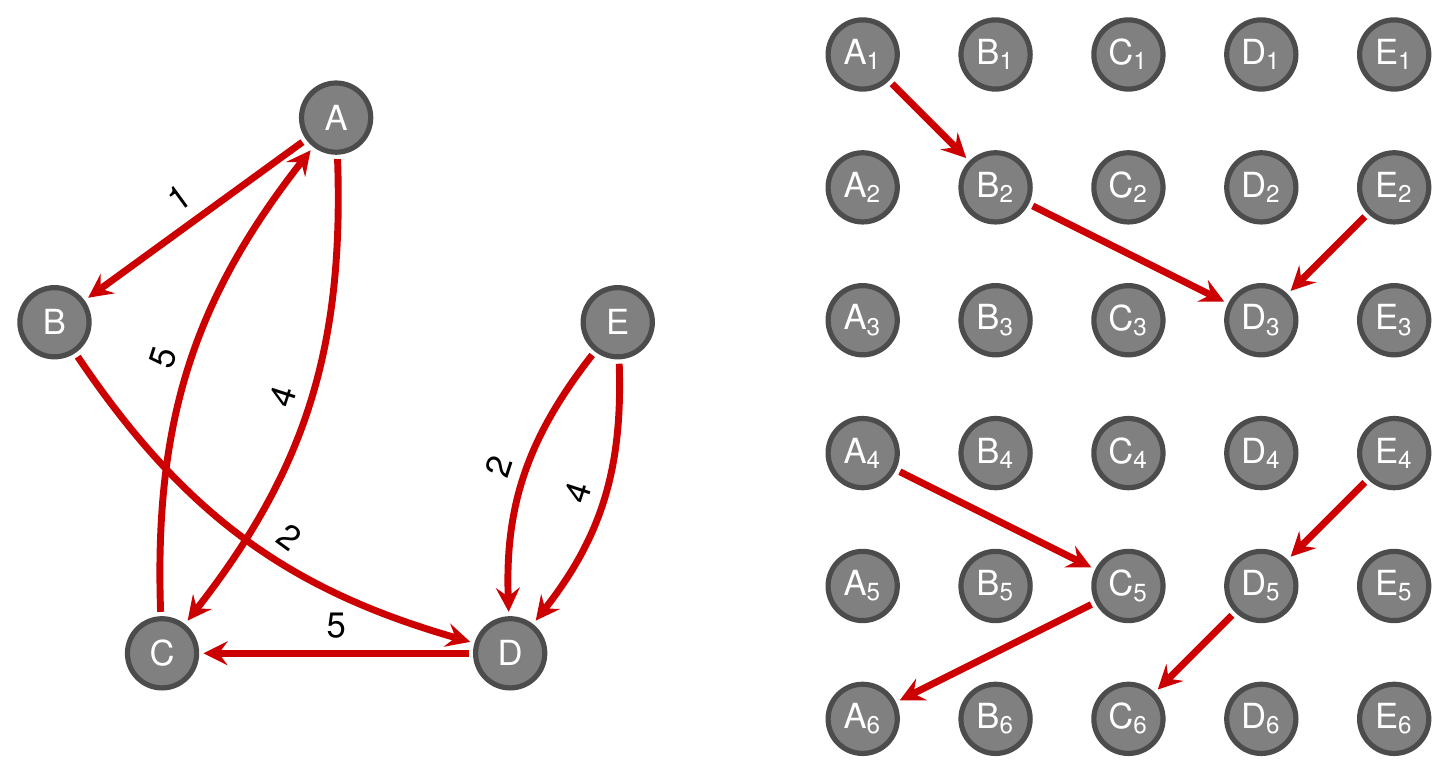}
	\caption{
		A example temporal network (left) and the corresponding time-unfolded model of the network (right).
		Edges in the temporal network at labelled in with the time at which they appear (in this case, instantaneously).
		In the time-unfolded model nodes are replaced by time-indexed copies of themselves.
		Edges are then created between a time-index node and another time-indexed node that is strictly further ahead in time.
		Nodes can be connected to nodes in the next time interval for a discrete time model of instantaneous contact (as presented here), or to nodes at some later point in time for contacts with durations. 
	}
	\label{fig:time_unfolded_model}
\end{figure}

Where temporal networks differ most from their static counterparts is in path-counting.
Figure~\ref{fig:time_unfolded_model}(left) shows an example temporal network where edges are labelled with the time at which they occur.
If we consider only the induced aggregated static network we would observe a path from node E to node B, however no such path exists in the real network.
By simply ignoring the ordering of events we potentially exaggerate the number of possible paths through the network.
Naturally this can lead to overestimates when considering the proliferation of a spreading dynamic on the network.

One way to preserve the temporal ordering of paths is by considering time-unfolded (or time-node) graph~\cite{michail2016introduction, takaguchi2016coverage}.
Here nodes are replaced by time-indexed copies of themselves and edges can only travel from older to newer time-indexed nodes.
The utility of these graphs are that they are amenable to traditional static network methods and can be efficient to work with as they are both directed and acyclic.
Any analysis of the time-index nodes does need to be mapped back to the original nodes however this is often trivial.
Figure~\ref{fig:time_unfolded_model}(right) shows the corresponding time-unfolded model for the temporal network.
Using this model it is easy to confirm that there is no temporal path from E to B.

\subsubsection*{Higher-order Graphical Models}

The concept of a second-order model has been around for a long time in the form of line graphs~\cite{hemminger1978line}, De Bruijin graphs~\cite{de1946combinatorial}, or Hashimoto graphs~\cite{hashimoto1989zeta}.
In this class of model the edges of the original static network become the nodes of the new model, and edges are connected to other edges if they have a node in common.
In the temporal network literature these models have been referred to as \emph{memory networks}~\cite{lambiotte2014effect,rosvall2014memory}, owing to their use of modelling second-order Markov random walks, or random walks with \emph{memory}.
Models which incorporate memory are particularly useful for considering walks on temporal networks.
As an example, consider the temporal network of individual journeys across the London tube network.
These journeys are the result of many commuters travelling to and from work and can not be well modelled by a random walk on the underlying static network \cite{lambiotte2018understanding}.
Upon arrival at a particular station the likelihood of your next destination depends on, at the very minimum, of where you were last, e.g. you would be very unlikely to back-track unless you were particularly lost!
This type of network can however be captured with a higher-order model.

\begin{figure}[!h]
	\centering
	\includegraphics[width=0.7\linewidth]{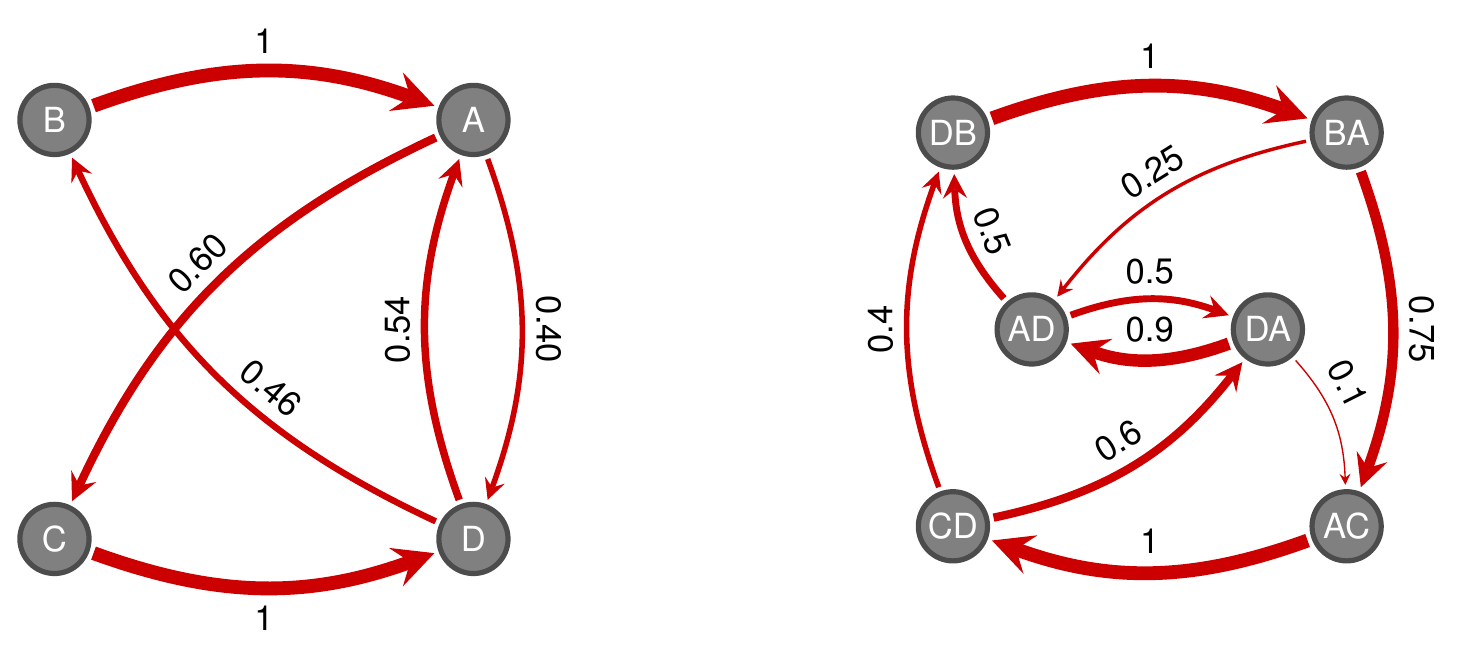}
	\caption{
		An example of a first-order (left), and second-order (right) graphical model.
		In each case a random walker can transition between states with some probability (indicated by the edge weights).
		In the first-order model, the probability of transition between nodes is conditional only on the current node whereas in the second-order model the transition is dependent on both the current and previous node.
		For example, $P(X_{t+1}=D|X_t=A, X_{t-1}=Z) = 0.4$ for any node $Z$, whereas $P(X_{t+1}=D|X_t=A, X_{t-1}=B) = 0.25$ and $P(X_{t+1}=D|X_t=A, X_{t-1}=D) = 0.9$.
	}
	\label{fig:higher_order_model}
\end{figure}

Figure~\ref{fig:higher_order_model} shows an example of a first- and second-order model for a random walk process on a network.
Edge weights represent the probability of traversing from state to state.
In the first-order model the likelihood of the traversal of an edge is dependent only on the current node.
In contrast the probability of traversing the D $\to$ B edge in the second-order model depends whether we have previously arrived at D from C or A (probabilities $0.4$ and $0.5$ respectively).

Beyond second-order Markov models there have been recent works which consider more general higher- or variable-order models~\cite{scholtes2016higher,scholtes2017network}.
Unfortunately the number of possible states increases exponentially with the chosen model order which can lead to computational intractability.
This issue can however be remedied to some extent by considering higher-order states as and when a lower-order representation is insufficient to describe the data~\cite{xu2016representing}.

Finally it is worth remarking that `higher-order' interaction has a secondary meaning beyond higher-order Markov models. 
The models outlined above have all been based on the assumption of pairwise, or dyadic, interaction.
However, much like the static network assumption, the idea of dyadic interaction is often more convenient than it is valid.
The concept of hypergraphs~\cite{berge1984hypergraphs}, where interactions (edges) can include multiple nodes, and the corresponding directed hypergraphs~\cite{gallo1993directed} have been around for multiple decades and have been well studied.
More recently however their has been work studying non-dyadic interaction over time through both topological perspectives, in simplicies and simplicial complexes~\cite{benson2018simplicial}, or by considering the structure of a growing and evolving hypergraph~\cite{taramasco2010academic}.

\subsubsection*{Contributions}

We generalise the concept of an event graph and show that event graphs can be seen as a family of second-order models for temporal data, characterised by an event joining function.
We also show that this provides a way to model non-dyadic interactions between nodes with minimal effort.
To the author's knowledge this is the first model for sequences of non-dyadic events at a network-level (sequences of events have been modelled at an individual level however~\cite{benson2018sequences}).
To illustrate the event graph model we provide a number of examples of its application to a selection of temporal networks.
These examples cover previous research and more prospective examples which require further study.

In Section~\ref{sec:event_graph} we introduce the event graph and in Section~\ref{sec:hypergraphs} we show how it can be can be generalised to include non-dyadic interaction in the form of hyper-events.
In Section~\ref{sec:applications} we give a number of example applications of the event graph (both novel examples and examples taken from previous works) before concluding in Section~\ref{sec:conclusion}.

\section{Event Graphs}
\label{sec:event_graph}

Our definition of an event graph combines both the notion of a higher-order model (in particular a \emph{second-order} model) with that of a time-unfolded model.
We consider a temporal network defined by a sequence of temporal events $(e_i)_{i=1}^M$ where each event is a triplet of the form $e_i = (u_i,v_i,t_i)$ (as described in Section~\ref{sec:introduction}).
For the purpose of generality we remain ambiguous as to whether this represents a directed event $u_i \to v_i$ or an undirected event $u_i \leftrightarrow v_i$, although we clarify when needed.
We also make the assumption that a node may participate in only one event at a time, although as we see in Section~\ref{sec:hypergraphs} this restriction can managed by considering \emph{hyper-events}.

For any two events we can examine the number of shared nodes, and the time between the two events occurring, i.e. the \emph{inter-event time}.
\begin{definition}[Inter-event Time (IET)]
The inter-event time $\tau$ between two events $e_i=(u_i,v_i,t_i)$ and ${e_j=(u_j,v_j,t_j)}$ is given by
\begin{align*}
	\tau(e_i,e_j) = \begin{cases}
		t_j-t_i & \text{ if } t_j > t_i \\
		0 & \text{ otherwise}.
	\end{cases}
\end{align*}
If the events are not instantaneous and have durations $\delta_i$ and $\delta_j$ respectively then
\begin{align*}
	\tau(e_i,e_j) = \begin{cases}
		t_j-(t_i+\delta_i) & \text{ if } t_j > t_i + \delta_i \\
		0       & \text{ otherwise}.
	\end{cases}
\end{align*}
\end{definition}
For simplicity of notation we write $\tau_{ij} = \tau(e_i,e_j)$ from here on in.

We now define the event graph in full generality before describing particular examples.
\begin{definition}[Event Graph]
\label{def:event_graph}
An event graph $G$ is a directed static graph given by the tuple $G = (E, f_E)$ where $E$ is a set of temporal events, and $f_E: E \times E \to [0,1]$ is a binary function which prescribes the edges of the graph. 
If $f_E$ has no explicit dependence on the set of events then it is denoted $f$.  
\end{definition}
Furthermore, it is often useful to consider a weighted event graph $G^\tau$ which instead uses a joining function ${f^\tau: E \times E \to \mathbb{R}_0^+}$ which is weighted by the IET, i.e. $f^\tau(e_i,e_j) = \tau_{ij} f (e_i, e_j)$.
The weighted event graph is topologically identical to the event graph however the extra information in the edge weights provides a means to threshold edges and investigate the percolation properties (see Section~\ref{sec:applications}).

\subsection{Joining Functions}

\begin{figure}[!h]
	\centering
	\begin{subfigure}{0.35\linewidth}
		\centering
		\includegraphics[width=\linewidth]{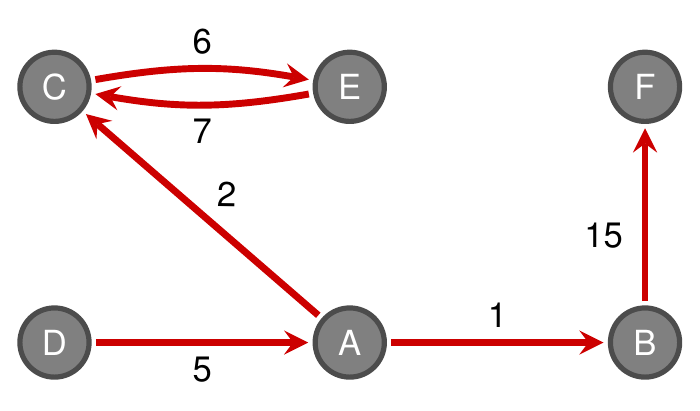}
		\caption{}
	\end{subfigure}
	\vspace{0.0em}

	\begin{subfigure}{0.23\linewidth}
		\centering
		\includegraphics[width=\linewidth]{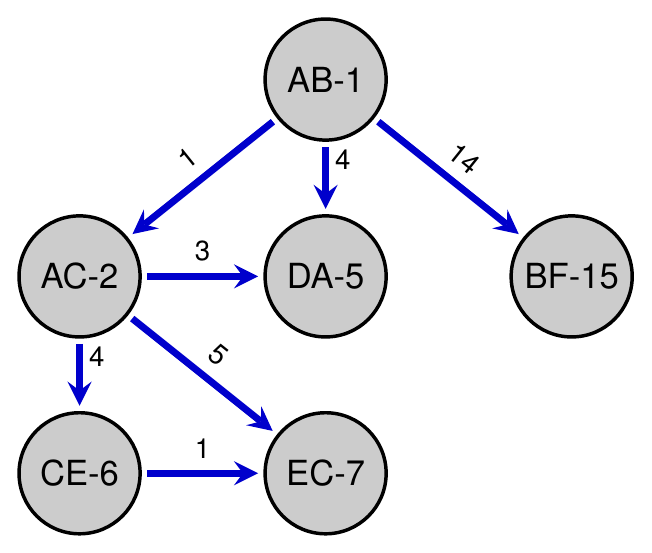}
		\caption{}
	\end{subfigure}
	\begin{subfigure}{0.23\linewidth}
		\centering
		\includegraphics[width=\linewidth]{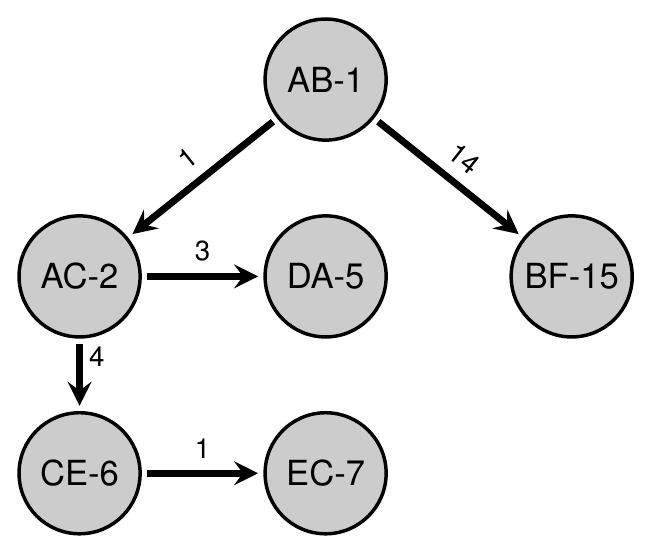}
		\caption{}
	\end{subfigure}
	\begin{subfigure}{0.23\linewidth}
		\centering
		\includegraphics[width=\linewidth]{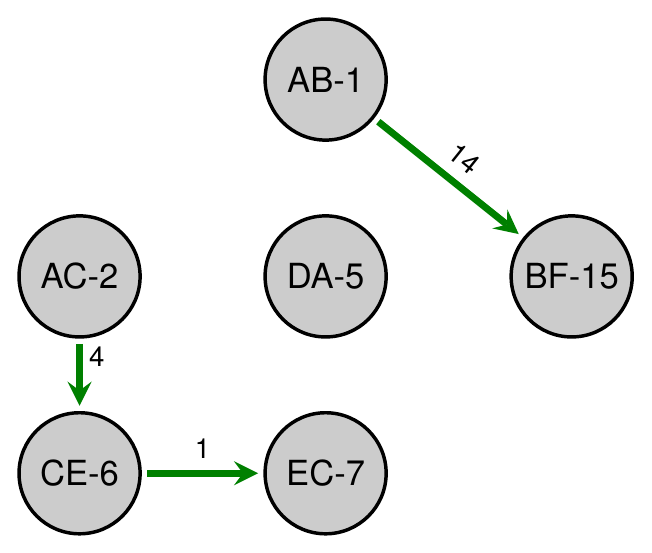}
		\caption{}
	\end{subfigure}
	\begin{subfigure}{0.23\linewidth}
		\centering
		\includegraphics[width=\linewidth]{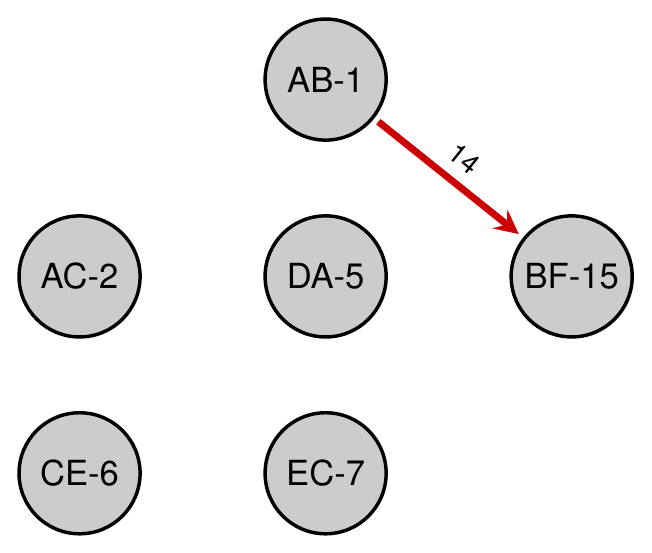}
		\caption{}
	\end{subfigure}
	\caption{
		Event graph examples using multiple joining rules. 
		(a) An example temporal network. 
		Edges are labelled with the time at which they occur.
		(b) The $\Delta t$-adjacent event graph with $\Delta t = 20$.
		(c) The node-subsequent $\Delta t$-adjacent event graph with $\Delta t = 20$.
		(d) The walk-forming event graph (with no $\Delta t$ restriction).
		The prescription of path-forming events is much stricter than $\Delta t$-adjacency hence there are far fewer edges in the event graph.
		(e) The minimum-gap non-backtracking event graph with ${(\Delta t_1, \Delta t_2) = (5,15)}$.
		Events now have a minimum inter-event time required for them to be connected resulting in fewer edges.
		}
	\label{fig:event_graph_examples}
\end{figure}

The event graph describes a family of different graphs, modulated by the joining function $f_E$.
This has parallels with the family of graphs in a preferential attachment model generated with different attachment kernels~\cite{krapivsky2001organization}.
In a similar fashion, the choice of joining function has a major effect on the structure of the event graph.
The choice of joining function should be dependent on the nature of the temporal network or the dynamic under study (contagion, social interaction, etc.).
Furthermore, the computational cost of constructing an event graph varies strongly with the choice of joining rule so this should also be considered.
The remainder of this section is devoted to exploring particular joining functions which have been recently been proposed and used previously.
In Figure~\ref{fig:event_graph_examples} we give an example temporal network and the resulting event graphs for different joining rules.

\subsubsection*{$\bm{\Delta t}$-Adjacency}

One joining rule (and perhaps the simplest) is $\Delta t$-adjacency, first defined in~\cite{kovanen2011temporal}.
Following their definition, two events are $\Delta t$-adjacent if they share at least one node, and the time between the two events (IET) is no greater than $\Delta t$, for some prescribed $\Delta t \in \mathbb{R}^+$.
Furthermore, two events are $\Delta t$-connected if they can be joined by a sequence of pairwise $\Delta t$-adjacent events.
As a function, $\Delta t$-adjacency can be written as
\begin{align*}
f(e_i,e_j) = (0 < \tau_{ij} \leq \Delta t) \land \left( \{u_i, v_i\} \cap \{u_j,v_j\} \neq \emptyset \right)
\end{align*}
which simply states that the intersection between node sets must be non-empty.
Adjacency makes intuitive sense assuming that there are no external (or unobserved) interactions between agents; information can only be transmitted between two events if there is one or more common nodes where that information can persist.
The parameter $\Delta t$ is an upper bound for how long this information could persist, or how close two events need to be for us to consider a causal relationship.
In the limit $\Delta t \to \infty$ there are no restrictions on temporal proximity, and the requirement is only that events are adjacent.

An event can be connected to any number of subsequent events provided the adjacency criteria is met. 
However, in certain cases we are interested in only the subsequent adjacent event for a given event, or the subsequent event for each node in the given event (as in~\cite{kovanen2011temporal, mellor2017temporal}).
Let
\begin{align*}
	A^t(S) = \left\{(u,v,t') \in E \text{ s.t. } S \cap \{u,v\} \neq \emptyset \text{ and } t'>t \right\}
\end{align*}
be the set of all events that a set of nodes $S$ participates in after a time $t$.
These joining rules can then be represented by
\begin{align*}
	f_E(e_i,e_j) = (0 < \tau_{ij} \leq \Delta t) \land \left( j= \min\{k|e_k \in A^{t_i}(\{u_i,v_i\}\} \right).	
\end{align*}
and 
\begin{align*}
	f_E(e_i,e_j) = (0 < \tau_{ij} \leq \Delta t) \land \left( \bigvee_{S \in \{\{u_i\},\{v_i\}\}} (j= \min\{k|e_k \in A^{t_i}_s\}) \right).
\end{align*}
for subsequent event and subsequent event for each node respectively.
For convenience we will refer to graphs created these extra constraints as \emph{event-subsequent} and \emph{node-subsequent} $\Delta t$-adjacent event graphs.
Note that this is an example where we require knowledge of the set of all events (to ensure we pick the subsequent one).
However if events are added to the graph sequentially then these joins can be made independently of the event set since any future events cannot be subsequent by definition.

\subsubsection*{Walk-forming}

Temporal paths and walks\footnote{
	Temporal paths are distinguished from temporal walks by the number of times the walk/path can visit each node. 
	Temporal paths may only visit each node at most once, whereas temporal walks can visit each node multiple times.
} are an important feature of temporal networks as, analogously to paths and walks in static networks, they allow for the dissemination of information through the network.
Furthermore understanding the number of possible walks and paths between nodes is crucial for controllability and route-finding.

For undirected networks any two events that are adjacent are walk-forming, however for directed networks further distinction is needed.
In this case a suitable joining rule is
\begin{align*}
f(e_i,e_j) = (0 < \tau_{ij} \leq \Delta t) \land ( u_j = v_i ),
\end{align*}
which states that the target of the first event must be the source of the second event.

\subsubsection*{Minimum Gap and Non-backtracking}

For certain temporal network processes it may be suitable to introduce a minimum time between events occurring for them to be considered to be connected. 
Once such example is for aviation networks, i.e. the network of scheduled flights, where it is an unrealistic assumption that a connection can be made at an airport without first taking time to traverse across the airport from one plane to another~\cite{gao2017schedule}.
This motivates us to introduce a non-zero lower bound on $\tau_{ij}$.
Keeping with the air travel example, it would also be pointless to immediately return back along the same route that you have just travelled so it would be sensible to remove these backtracking possibilities.
Together these rules combine to give
\begin{align*}
f(e_i,e_j) = (\Delta t_1 < \tau_{ij} \leq \Delta t_2) \land ( u_j = v_i ) \land ( v_j \neq u_i ),
\end{align*}
where $\Delta t_1, \Delta t_2 \in \mathbb{R}^+$ and $\Delta t_1 < \Delta t_2$.

The non-backtracking event graph draws some similarity to the non-backtracking matrix representation of a static graph~\cite{horton2006zeta}, or the Hashimoto matrix~\cite{hashimoto1989zeta}.
Non-backtracking operators have previously been used in the static case for community detection~\cite{krzakala2013spectral} and spectral clustering however their extension to the temporal setting has yet to be explored.

\section{Non-dyadic Interactions}
\label{sec:hypergraphs}

Although many studies of networks (and temporal networks) revolve around the assumption of pairwise interaction between nodes, this assumption often does not hold or is an oversimplification.
In networks of academic collaboration, cliques are formed between authors when they publish together.
These interactions are non-dyadic, although when abstracting to a network an individual edge between two authors is considered independently of the clique it came from (for publications with three or more authors).
Email correspondence and other digital communication is also another area where non-dyadic interaction occurs.
Emails can be sent to any number of people in a single interaction however this type of behaviour is indistinguishable from multiple person-to-person emails if interactions are aggregated over time.
These over simplifications have peculiar consequences when we perform further analysis on these networks such as clustering or path-counting.

Fortunately these higher-order interactions can be captured within an event graph by introducing the notion of \emph{hyper-events} and the \emph{temporal hypergraph}.
\begin{definition}[Temporal Hypergraph]
	Let $V \subseteq \mathbb{N}$ be a set of nodes, $T \subseteq \mathbb{R}^+_0$ a set of times, $D \subseteq \mathbb{R}^+_0$ a set of durations, and $E$ a set of temporal hyper-events.
	A temporal hypergraph is defined by the quadruple $G = (V,T,D,E)$ where temporal hyper-events take the form
	\begin{align*}
		e_i = (U_i, t_i, \delta_i)
	\end{align*}
	where $U_i \subseteq V$.
	For \emph{directed} temporal hypergraphs hyper-events take the form
	\begin{align*}
		e_i = (U_i, V_i, t_i, \delta_i)
	\end{align*}
	where $U_i,V_i \subseteq V$ and $U_i \cap V_i = \emptyset$ analogously to non-temporal hypergraphs.
\end{definition}
If we fix $|U_i|=k$ for all $i$ we call the resulting undirected hypergraph $k$-\emph{uniform}.
Similarly, for directed events if we restrict $|U_i|=k$ and $|V_i|=m$ for all $i$ we have a $(k,m)$-\emph{uniform} hypergraph.

This generalises the concept of an event sequence, and hence extends temporal networks to include non-dyadic interactions.
If we consider a $2$-uniform (or $(1,1)$-uniform) hypergraph we see that we arrive back at the familiar undirected (and directed) event sequences of the previous section.
As before we make the assumption that a node may be involved with at most one event at any given time, although now a node may have multiple edges during the event.
The definition of the event graph for temporal hypergraphs is identical to the definition for dyadic interaction (Definition~\ref{def:event_graph}).

\subsection{Joining Rules}

While the event graph definition is unchanged for non-dyadic interaction the joining rules for hyper-events require further generalisation.

\begin{figure}[!h]
	\centering
	\begin{subfigure}{0.45\linewidth}
		\centering
		\vspace{.61em}
		\includegraphics[width=\linewidth]{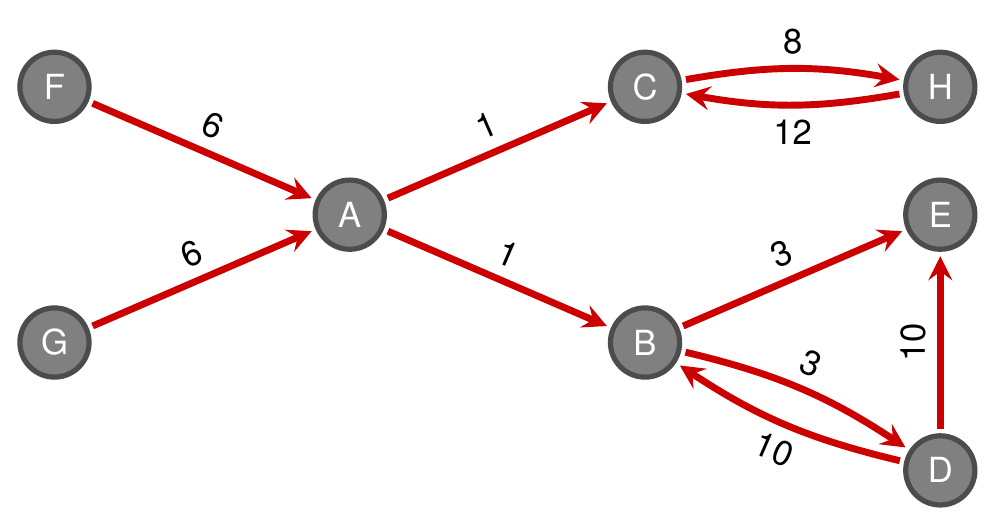}
		\vspace{.61em}
		\caption{}
	\end{subfigure}
	\hspace{2em}
	\begin{subfigure}{0.31\linewidth}
		\centering
		\includegraphics[width=\linewidth]{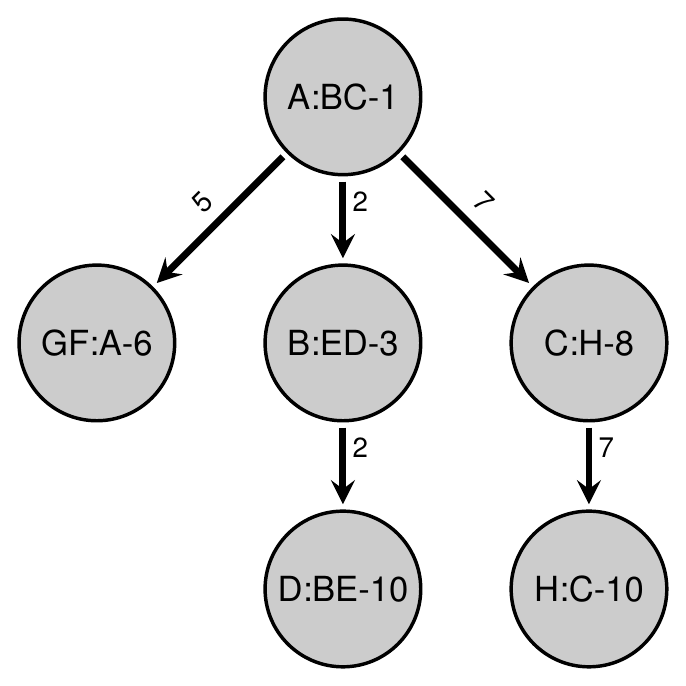}
		\caption{}
	\end{subfigure}
	\caption{
		Event graph construction for temporal hypergraphs.
		In (a) we have an example temporal hypergraph where an event can include more than two nodes, represented here by the identical timestamp on ingoing or outgoing edges.
		In (b) we have the corresponding node-subsequent $\Delta t$-adjacent event graph.
		Events are labelled as \texttt{[Source Nodes]:[Target Nodes]-[Time]} and edges are weighted by the IET as before.
	}
	\label{fig:hyper_event_graph_examples}
\end{figure}

$\bm{\Delta t}$\textbf{-Adjacency.} For two hyper-events to be $\Delta t$-adjacent requires that one or more of the nodes in the first event (source or target) appear in the second event.
For directed events this equates to a joining rule
\begin{align*}
	f(e_i,e_j) = (0 < \tau_{ij} \le \Delta t) \land ( (U_1 \cup V_1) \cap (U_2 \cup V_2) \neq \emptyset).
\end{align*}
For undirected events this requirement is $U_1 \cap U_2  \neq \emptyset$.
Node-subsequent and event-subsequent adjacency can also be trivially defined for hyper-events.

\textbf{Walk-forming.} Walk-forming carries the stricter restriction of $(V_1 \cap U_2 \neq \emptyset)$ compared to $\Delta t$-adjacency. 
For undirected hyper-events all $\Delta t$-adjacent events are walk-forming.

\textbf{Minimum Gap and Non-backtracking.} The minimum gap condition is independent of the number of nodes in each event and is therefore unchanged.
The concept of `non-backtracking' is ill-defined for hypergraphs however.
The analogous restriction would be that $(U_1 \cap V_2 = \emptyset)$, however this is ignorant of instances where an event causes both backtracking and non-backtracking walks.

\section{Applications}
\label{sec:applications}

In this section we explore some of the analysis of temporal networks that has been conducted with event graphs, and highlight new avenues for further research through examples.
This is not an exhaustive list, however it should hopefully illuminate the possible applications of these higher-order models.

\begin{table}[!h]
	\centering
	\resizebox{\columnwidth}{!}{%
	\begin{tabular}{lcccrrp{7cm}} \toprule[1pt]
	\textbf{Name} & \textbf{Nodes} & \textbf{Events} & \textbf{Duration} & \textbf{Hyp.} & \textbf{Dir.} & \textbf{Description} \\ \midrule[0.5pt]
	\texttt{twitter-emirates}\cite{mellor2018classifying} & 53,251 & 167,664 & 1 day & Yes & Yes & Posts taken from Twitter using the keyword `Emirates.'\\
	\texttt{random-complete} & 500 & 50,000 & 50,163 & No & Yes & A random network where at each iteration a random node is connected to another random node.\\
	\texttt{academic-coauthors}\cite{salnikov2018cooccurrence} & 30,927  & 54,177  & 29 years & Yes & No & A network of article coauthorship taken from ArXiv (\texttt{math} and \texttt{math-ph}). \\
	\texttt{social-ucirvine}\cite{panzarasa2009patterns} & 1,899 & 59,835 & 193 days & No & Yes & An online social network from UC Irvine. \\
	\texttt{sociopatterns-primary}\cite{stehle2011high, url_sociopatterns} & 242 & 38,923 & 2 days & Yes & No & A proximity network of school children. \\ \bottomrule[1pt]
	\end{tabular}
	}
	\vspace{0.5em}
	\caption{Data descriptions.}
	\label{tab:data}
\end{table}

Throughout this section we will use a number of datasets (both real and synthetic) which are detailed in Table~\ref{tab:data}.
These cover a range of durations (from 1 day to 29 years), and varying numbers of nodes and events.
We have also included a mixture of directed and undirected networks as well as examples using hyper-events.
This data is readily available online at the respective references, aside from~\cite{salnikov2018cooccurrence} which is available upon reasonable request.
The \texttt{sociopatterns-primary} network in its original form is dyadic.
We generate a sequence of hyper-events by first aggregating all nodes which are connected (not necessarily pairwise connected) at the same time, and then check for the persistence of this set across time to create an event duration.
A new event is therefore created each time a node enters or leaves the node set.
The \texttt{random-complete} network is generated synthetically in an iterative fashion.
At each timestep, two nodes are picked at random with one chosen randomly to be the source, and the other the target.
Time is advanced by a random increment drawn from an exponential distribution with mean one.
This is equivalent to randomly sampling edges from a complete graph.

One final point to note is that these methods conducted on the event graph are agnostic to whether the events are directed/undirected or dyadic/non-dyadic\footnote{
	Some methods require further adaptation to be applicable to non-dyadic interaction such as centrality calculations (Section~\ref{subsec:centrality}).
}.

\subsection{Percolation}

This aspect of event graphs has recently received the most focus in~\cite{kovanen2011temporal, kivela2018mapping} and~\cite{mellor2017temporal}.
These examples use $\Delta t$-adjacency as the joining rule for the event graph.
The introduction of the parameter $\Delta t$ naturally poses the question of how the structure of the event graph varies with $\Delta t$.
The answer to this question is non-trivial as the event graph incorporates not only the distribution of the inter-events times, but is also a function of the topological interactions between nodes (i.e. the underlying graph structure).

Provided the joining rule prohibits events to be connected to events in the present or past the event graph is a directed acyclic graph.
We therefore consider the weakly connected components of the event graph as there are no strongly connected components, and adopt the shorthand name of temporal components\footnote{
	For the case of $\Delta t$-adjacency these components are also referred to as `maximal $\Delta t$-connected subgraphs'~\cite{kovanen2011temporal}.
}.
Let $C_{\Delta t}$ be the set of temporal components in the $\Delta t$-adjacent event graph, and let each component $c$ be defined by the event set $E^c \subseteq E$.
To illustrate the characteristics of the event graph we consider three properties of temporal components: the number of events, the number of nodes, and the component duration.
These are defined for directed networks by 
\begin{align*}
N_{\rm events}(E) &= |E|,\\
N_{\rm nodes}(E) &= \left| \bigcup_{(U,V,t) \in E} (U \cup V) \right|,\\
D(E) &= \max_{(U,V,t) \in E} t - \min_{(U,V,t) \in E} t,
\end{align*}
respectively.
Equivalent expressions can be given for undirected networks.

\begin{figure}[!h]
	\centering
	\begin{subfigure}{0.45\linewidth}
		\centering
		\includegraphics[width=\linewidth]{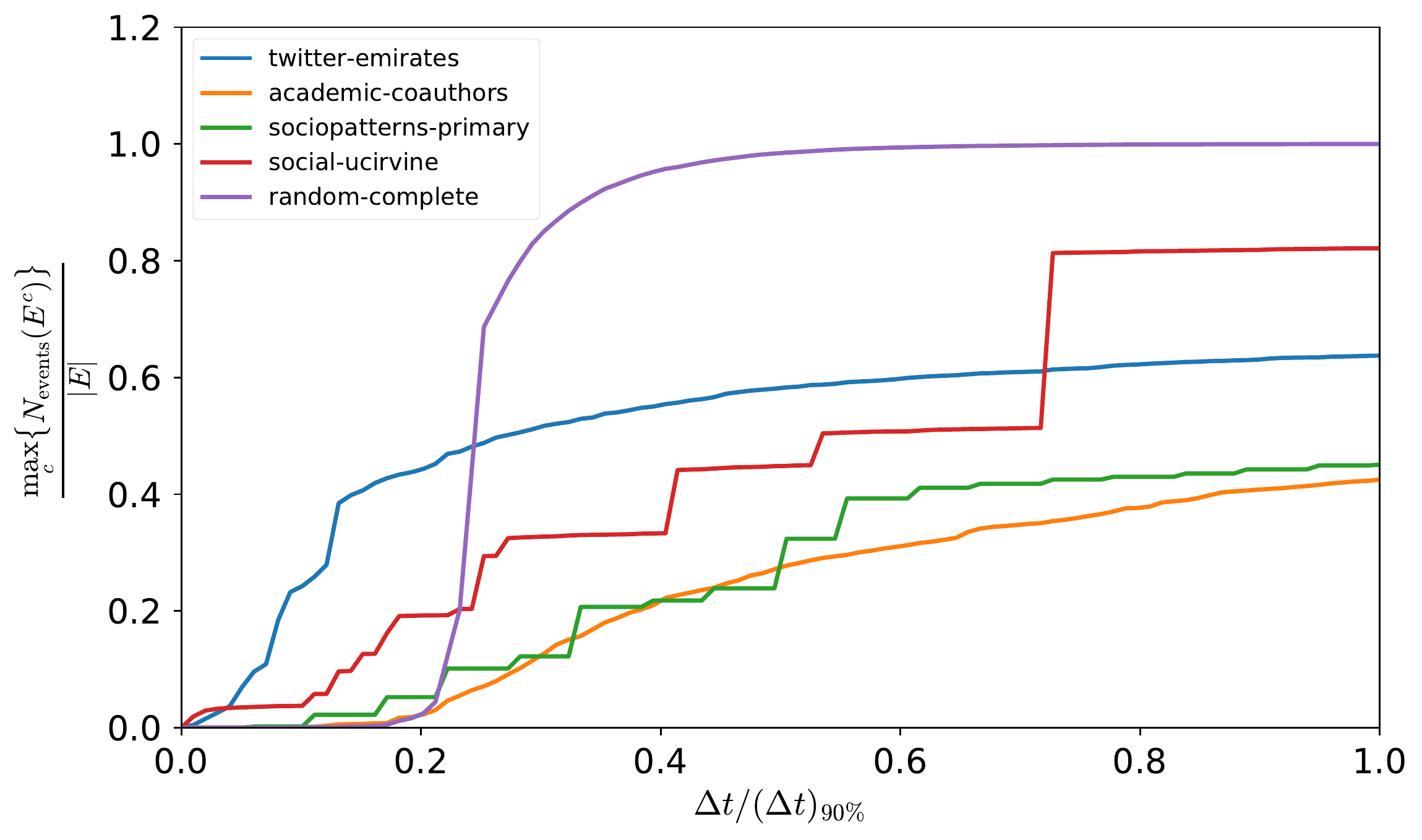}
		\caption{Events}
	\end{subfigure}
	\begin{subfigure}{0.45\linewidth}
		\centering
		\includegraphics[width=\linewidth]{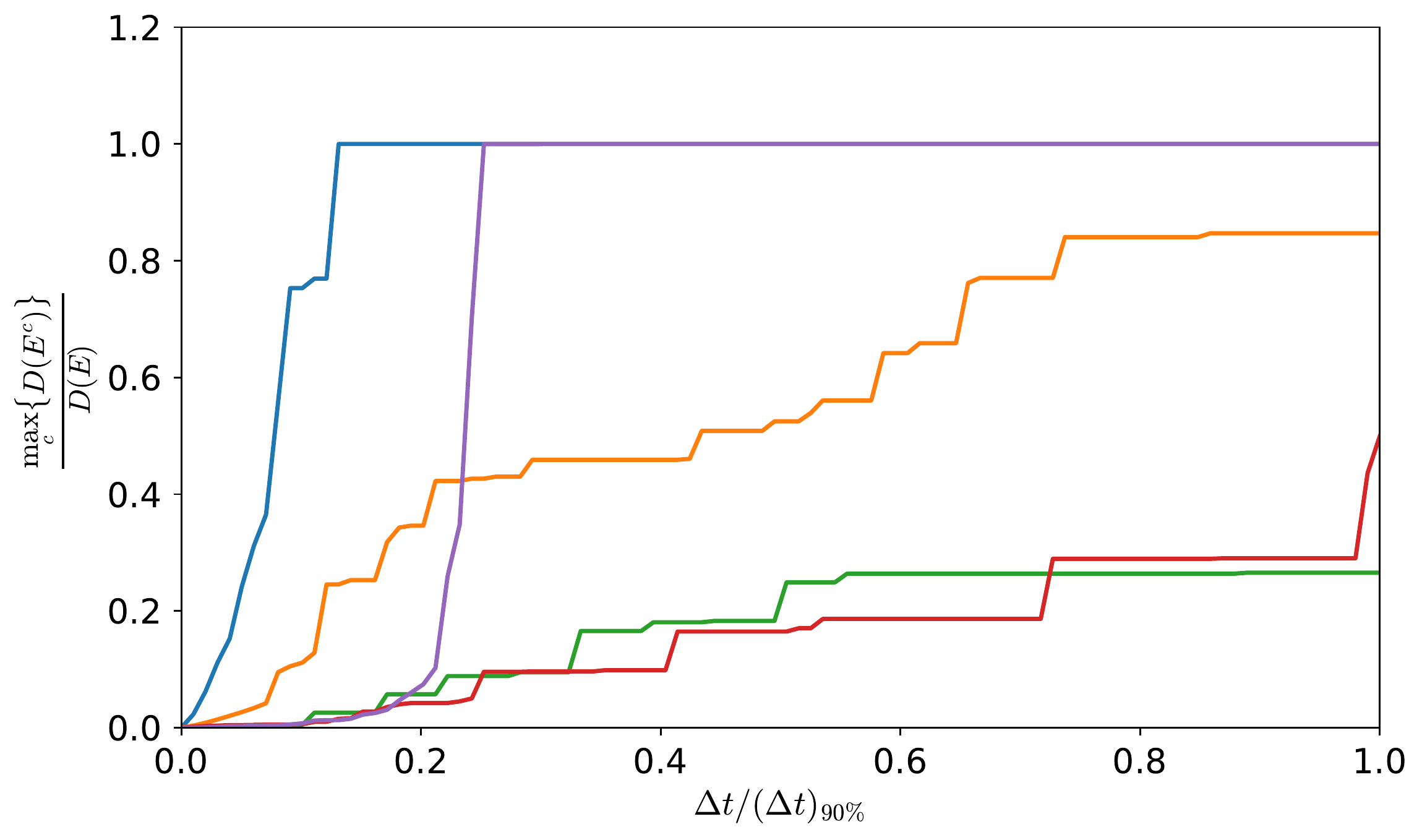}
		\caption{Duration}
	\end{subfigure}
	\vspace{1em}

	\begin{subfigure}{0.45\linewidth}
		\centering
		\includegraphics[width=\linewidth]{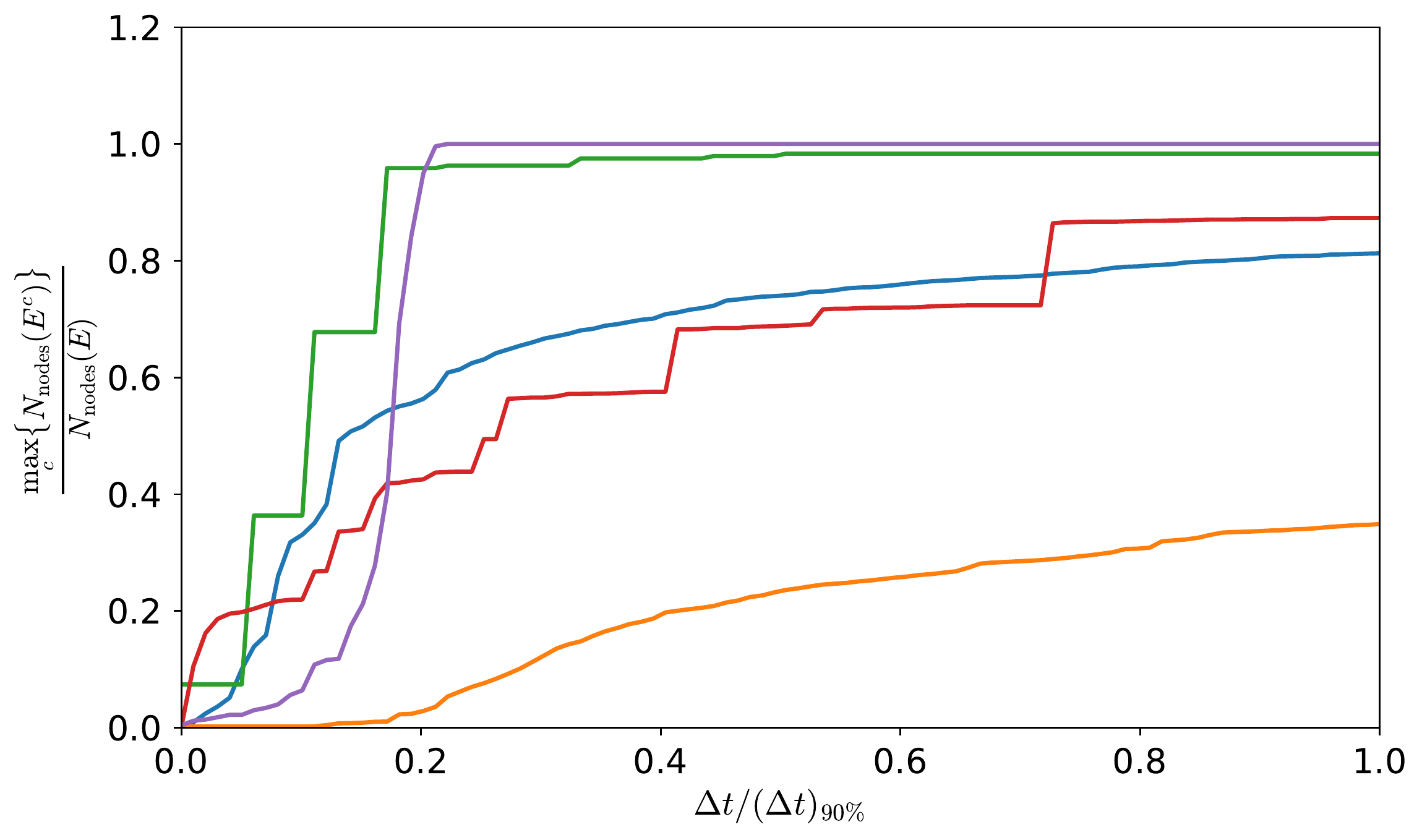}
		\caption{Nodes}
	\end{subfigure}
	\begin{subfigure}{0.45\linewidth}
		\centering
		\includegraphics[width=\linewidth]{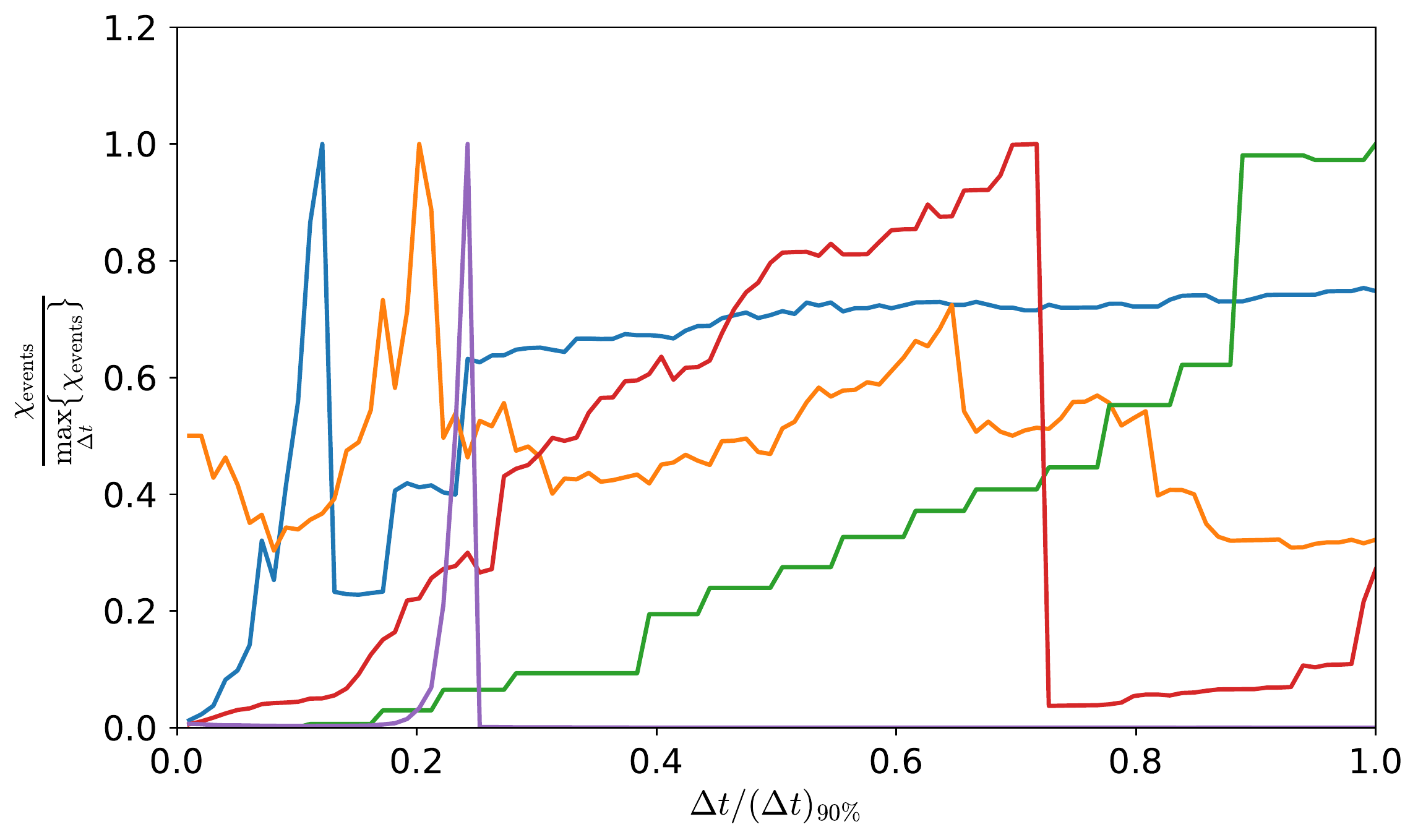}
		\caption{Event Susceptibility}
	\end{subfigure}
	\caption{
		The size of the largest component in the event graph in terms of (a) the number of events, (b) the duration of the component, and (c) the number of nodes in the component.
		Each is reported as a fraction of the total number of events, total duration, and total number of nodes of the data respectively.
		The event susceptibility is given in (d), defined as the average squared-size of all components barring the largest.
		To compare across datasets the susceptibility has been rescaled by the maximum value attained.
		The parameter $\Delta t$ is rescaled by the 90th percentile of the IET distribution for each dataset.
		}
	\label{fig:component_size_evolution}
\end{figure}

The evolution of the largest component sizes (for these three properties) for a range of values of $\Delta t$ is given in Figure~\ref{fig:component_size_evolution}(a)-(c).
To be able to compare the different networks we rescale $\Delta t$ by the 90th percentile of the IET distribution of the event graph.
The clear standout network is the \texttt{random-complete} network which displays a single transition across all three properties for a single value of $\Delta t$.
The \texttt{twitter-emirates} network makes up 60\% of events and 80\% of nodes as $\Delta t\to(\Delta t)_{90\%}$, however for a small value of $\Delta t$ there is a temporal component which spans the entire time period.
This is explainable by the presence of automated accounts (also known as \emph{bots}) which can be programmed to post messages periodically.
Once $\Delta t$ exceeds the periodicity of the bot the temporal component which contains the bot will have duration which spans the entire time period and would extend ad infinitum with more data collection.
The \texttt{sociopatterns-primary} network sees that most nodes feature in the largest component for small $\Delta t$.
As we discuss later, this is problematic when trying to prevent a spreading process from proliferating across the network.
Finally, the largest component in the \texttt{academic-coauthors} network has the smallest relative size for both nodes and events.
This is somewhat expected given that the effort required to jointly publish a paper together far exceeds that of sending a single message having a conversation.
Furthermore, while cross-group and cross-field collaborations may occur, they are more likely to be infrequent so these connections may well fall into the last decile of the IET distribution.

Recent work considers a toy model which samples events from an Erd\H{o}s-R\'{e}nyi network, and investigates the connectivity of the event graph as a function of the average degree of the underlying network, as well as $\Delta t$~\cite{kivela2018mapping}.
The particular focus of this work is to find the critical point $\Delta t_c$ which separates a disconnected event graph to that which contains a significant largest component.
In addition to the largest temporal component, they also investigate the average squared-size of all other components, excluding the largest.
This is referred to as the \emph{susceptibility} (denoted by $\chi$ and originating from magnetic susceptibility in statistical physics).
The peaks in susceptibility are associated with critical points or timescales in the network.
For their toy model (and similar to the \texttt{random-complete} network) there is one peak associated with the rapid increase in component size.
The picture for real-world networks however is not so clear. 
Investigating over a network of sexual-interaction, a mobile communications network, and a network of air transportation (flights), shows that although there may be a large dominant peak (or critical point), there are often other smaller peaks.
This hints at there being multiple timescales involved in the formation of the temporal network.
We can see examples of this in Figure~\ref{fig:component_size_evolution}(d).
The \texttt{random-complete} network has a clear peak, whereas for \texttt{twitter-emirates} and \texttt{academic-coauthors} there are multiple peaks.
For the \texttt{sociopatterns-primary} dataset such a peak does not exist, but instead exhibits a step-wise increase with $\Delta t$.

What does percolation analysis tell us about dynamics on these networks?
The $\Delta t$-adjacent event graph captures all temporal walks through the temporal network\footnote{
	It is also trivial to show that the components of the node-subsequent $\Delta t$-adjacent event graph are equivalent to those of the $\Delta t$-adjacent event graph.
} and hence captures how well a process can spread through the temporal network.
The parameter $\Delta t$ represents the minimum time that a spreading process must persist on a node in order for it to spread effectively.
Consider an epidemic model on the \texttt{sociopatterns-primary} dataset.
If the recovery time from infection is less than 0.2 (working in units relative to the 90th IET percentile) then the spreading process can only reach a fraction of all nodes under perfect conditions (Figure~\ref{fig:component_size_evolution}(c)), whereas if the recovery time is longer then it can potentially reach all nodes.
This type of information is incredibly useful and allows us to ascertain the possible reach of a spreading process starting at a particular point in time, or understand how to reduce spread by removing certain interactions of events (see Section~\ref{subsec:centrality}). 

\subsection{Motif Counting}

Motifs are crucial in helping to understand the structure and function of networks and have seen applications across a number of fields, in particular in the biological sciences~\cite{alon2007network, milo2002network, prvzulj2007biological}.
Motifs in temporal networks are however less well understood, and their is not a unified definition of a temporal motif.
Certain definitions consider compare the underlying induced static graph (ignoring temporal ordering)~\cite{zhao2010communication}, or use heuristics for motif counting~\cite{gurukar2015commit}.  
We give the definitions of two types of temporal motif which we name the \emph{windowed temporal motif}~\cite{paranjape2017motifs} and \emph{sequential temporal motif}~\cite{kovanen2011temporal} for clarity, and show that their definitions can be unified using the event graph.

\begin{definition}[Windowed Temporal Motif~\cite{paranjape2017motifs}]
A $k$-node, $l$-event, $\delta$-temporal motif is a sequence of $l$ time-ordered events $(e_1, \dots, e_l)$ such that $t_l - t_i \le \delta$ and the induced aggregate network has $k$ nodes.
\end{definition}
\begin{definition}[Sequential Temporal Motif~\cite{kovanen2011temporal}]
An $l$-event $\Delta t$-temporal motif is a graph of $l$ events where all pairs $(e_i,e_j)$ of events are $\Delta t$-connected, that is there exists a sequence of events $(e_n)_{n=1}^k$ such that $e_i=e_1$,\dots,$e_k=e_j$ and all pairs of consecutive events are $\Delta t$-adjacent.
Furthermore it is required that all events for a node are consecutive (i.e. no events omitted).
\end{definition}

These definitions differ in two ways.
Firstly as the name suggests, the windowed motif has to occur strictly within a time window of length $\delta$, whereas the sequential motif requires only that there are a chain of events whose inter-event time is no less than $\Delta t$.
This can result in motifs with a duration much greater than $\Delta t$.
Secondly the sequential motifs require that all events of each node in the motif are consecutive, that is, if a node is involved in events $e_i,e_j,e_k$ (with $t_i<t_j<t_k$) then events $e_i$ and $e_k$ cannot form a motif since $e_j$ is omitted. 
This makes the calculation of motif counts considerably easier, however potentially more susceptible to noise as a `noisy' intermediate event may obscure a `true' motif. 

\begin{figure}[!h]
	\centering
	\captionsetup[subfigure]{justification=centering}
		\begin{subfigure}{0.35\linewidth}
		\centering
		\vspace{.9em}
		\includegraphics[width=\linewidth]{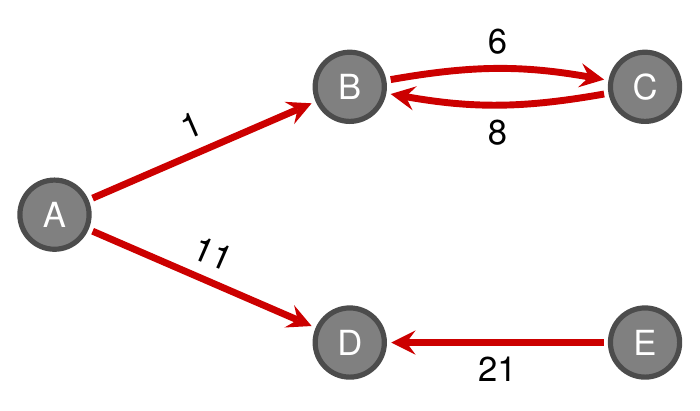}
		\vspace{.9em}
		\vfill
		\caption{}
	\end{subfigure}
	\begin{subfigure}{0.3\linewidth}
		\centering
		\includegraphics[width=\linewidth]{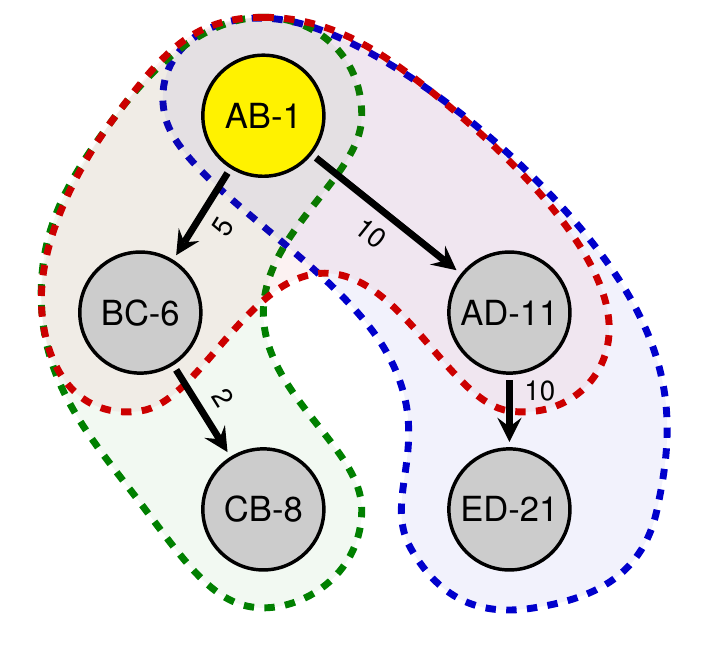}
		\caption{Sequential temporal motifs}
	\end{subfigure}
	\begin{subfigure}{0.3\linewidth}		
		\centering
		\includegraphics[width=\linewidth]{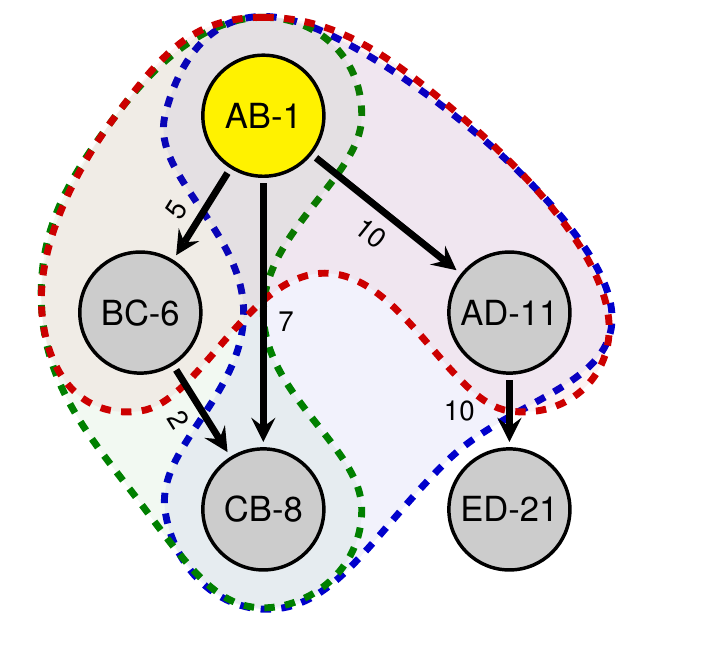}
		\caption{Windowed temporal motifs}
	\end{subfigure}
	\caption{
		Schematics for the calculation of temporal motifs using an event graph.
		In (a) we have an example temporal network, where edges are labelled at the time they occur.
		For (b) we construct the node-sequential $\Delta t$-adjacent event graph with $\Delta t = 10$.
		Sequential motifs (as described in~\cite{kovanen2011temporal}) can be found by either breadth-first or depth-first search from a node ensuring that any edge has weight of 10 or less.
		In this case there are three possible 3-event motifs which begin with the event A $\to$ B at time 1.
		For (c) we construct the  $\Delta t$-adjacent event graph with $\Delta t = 10$.
		We can then find windowed temporal motifs (as described in~\cite{paranjape2017motifs}) in the same fashion as (a), except this time the maximum path length cannot exceed 10.
		This also results in three 3-event motifs, however only two motifs are shared between definitions.
	}
	\label{fig:motif_calculation}
\end{figure}

Figure~\ref{fig:motif_calculation} shows how the motifs of a temporal network (a) can be represented as subgraphs of the node-subsequent $\Delta t$-adjacent event graph (b) and $\Delta t$-adjacent event graph
(c).
Starting from a given initial event (here highlighted in yellow), the associated three-event motifs can be found by traversing the edges of the event graph.
Sequential $\Delta t$-temporal motifs can be found by traversing edges whose weight is less than $\Delta t$ (or the event graph can be pre-pruned for efficiency).
Motifs generated in this fashion require a further check to confirm that all events for participant nodes are included if they fall within the timeframe of the motif.
This is referred to as a \emph{valid} temporal motif~\cite{kovanen2011temporal}.
Windowed $\delta$-temporal motifs require edges to be traversed ensuring that the maximum path length does not exceed $\delta$.
This is why (A,B,1),(A,D,11),(E,D,21) is not a valid windowed $\delta$-temporal motif with $\delta=10$.

Our above reasoning suggests that an alternative definition of temporal motifs (which can capture the different aspects) to be event graph subgraphs.
There are two choices to make, namely the choice of event graph joining rule and the subgraph search rule.
Either of these rules can be as tropical as required, depending on the use case.
This formalism also means that we can naturally extend motif counting to consider temporal hypergraphs.
Since the event graph is defined for hyper-events, temporal motifs are merely the subgraphs of these event graphs.
This does however further confound the issue of the number of possible motifs for a given number of events.
If the size of hyper-events are unrestricted then the results of any motif counting will be insignificant due to the low number of occurrences of each motif type.

We make no general comment on the computational complexity of finding temporal motifs, except that the problem is mapped to finding rooted subgraphs of a directed acyclic graph which satisfy certain constraints.
As the number of events in a motif increases, the number possible motifs increases exponentially which severely affects run-times.
More efficient algorithms for counting instances of a particular motif could also be devised by pruning the event graph for incompatible event interactions.

Lastly, one advantage this approach has over other purely counting algorithms~\cite{paranjape2017motifs} is that the motifs are enumerated and so their position in time is known.
This use useful for tracking the evolution of motif prevalence, as seen in Section~\ref{subsec:decomp}.
Furthermore, motifs can be assigned to individual nodes, enabling node-level counts of motifs which they participated in and also the role they play within that motif.

\subsection{Decomposition and Clustering}
\label{subsec:decomp}

\subsubsection{Decomposition}

Traditionally temporal networks have been decomposed into fixed-width intervals of time~\cite{holme2015modern, masuda2016guide}, typically a meaningful unit of time such as an hour, day, or week.
This allows a temporal network to be represented by a sequence of static adjacency matrices $(A_k)_{t=1}^T$ which themselves are amenable to traditional static network tools.
In discretising time into intervals we make the assumption that events within each interval are connected, and events between intervals are not.
However this leads to issues on the interval boundaries.
Suppose in a communication network a message is sent at 11:58pm and a message is returned at 00:02am the next day.
Intuition says that these two events are connected, however if time is discretised by day then these events are assumed to be unrelated.
Either event could however be connected to other events which are many hours away.
The fixed-width discretisation of time therefore sacrifices our beliefs about connectivity for computational and analytical ease.  

An event graph model offers an alternate approach, instead using the time between events as the decomposition threshold.
The decomposition of the temporal network is instead into the temporal components of the $\Delta t$-adjacent event graph.
This type of decomposition means that components can occur over a range of timescales, and components can overlap in time.

\begin{figure}[!h]
	\centering
	\begin{subfigure}{0.45\linewidth}
		\centering
		\includegraphics[width=\linewidth]{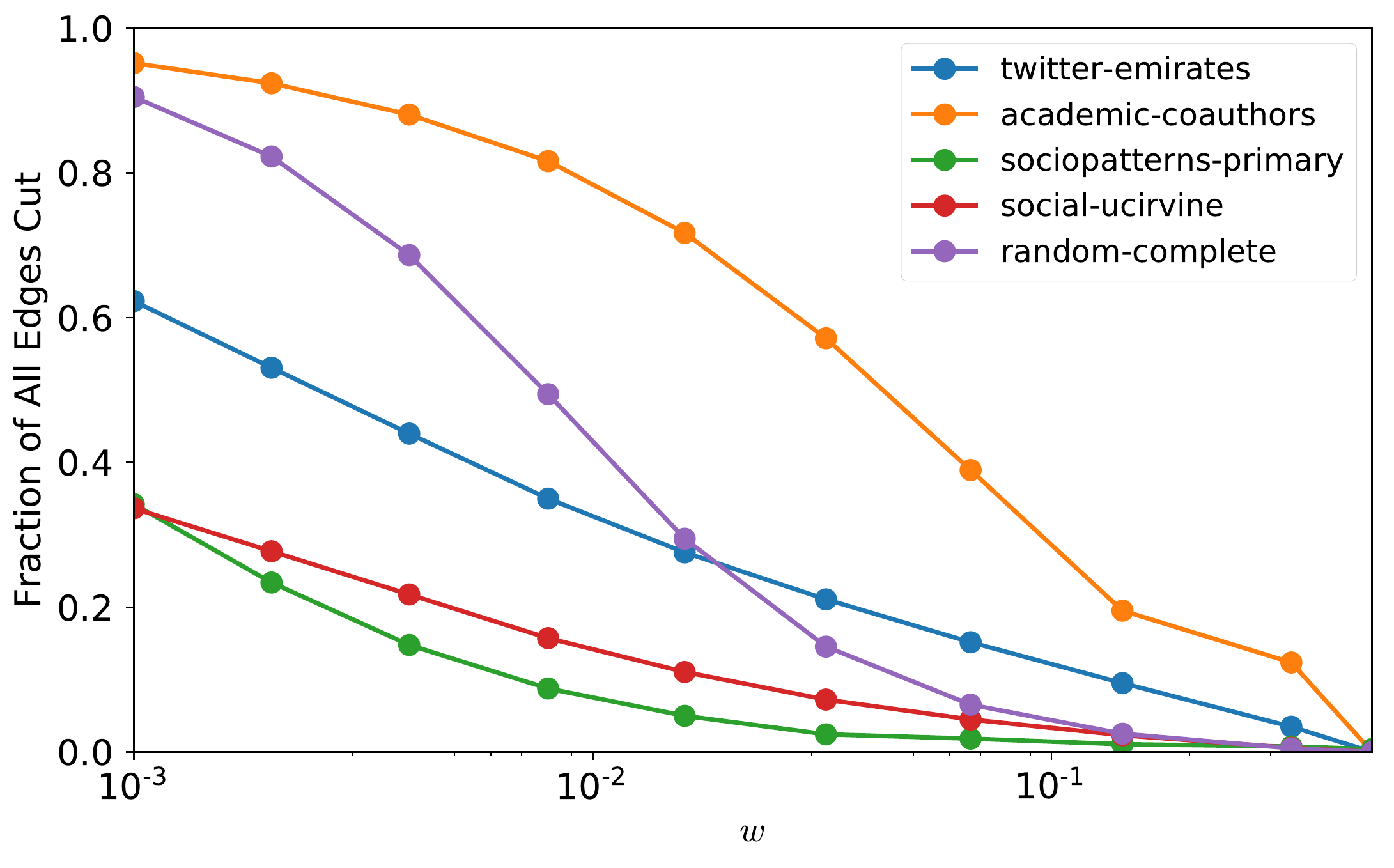}
		\caption{}
	\end{subfigure}
	\begin{subfigure}{0.45\linewidth}		
		\centering
		\includegraphics[width=\linewidth]{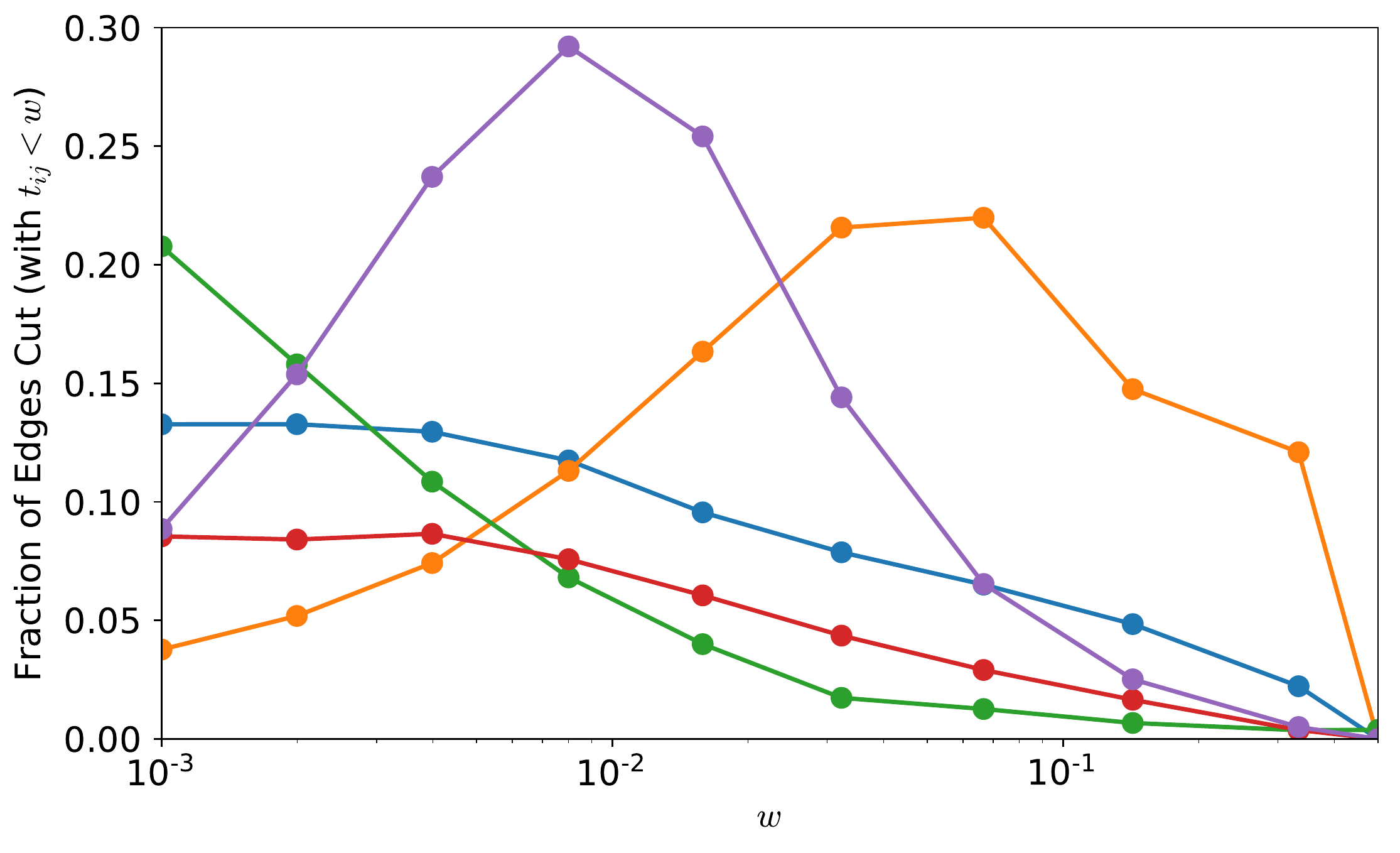}
		\caption{}
	\end{subfigure}
	\caption{
		Edge cutting in the event graph when considering fixed-width interval discretisation.
		(a) The fraction of edges cut in the node-subsequent $\Delta t$-adjacent event graph as a function of the interval width (given as a percentage of the total duration of the temporal network).
		As expected, the number of edges cut is a decreasing function of the interval width (smaller intervals cut potentially more interactions).
		There are however variable effects between datasets.
		(b) The fraction of edges cut whose weight (IET) is also less than the interval width.
		There are two datasets (\texttt{random-complete} and \texttt{academic-coauthors}) which show a non-monotonic relation with the interval width suggesting that an `intermediate' interval width is worse than either smaller or larger intervals.
	}
	\label{fig:percolation_cut}
\end{figure}

In Figure~\ref{fig:percolation_cut} we show the effect of fixed-width interval cutting.
Figure~\ref{fig:percolation_cut}(a) shows the fraction of edges cut in the node-subsequent adjacent
event graph (with $\Delta t \to \infty$) as a function of the interval width (expressed as a fraction of the total time period $w$.
Naturally this is a monotonically decreasing with $w$.
With the time period split into 1000, nearly all edges in the event graph are cut for \texttt{random-complete} and \texttt{academic-coauthors}.
The effect is smaller for the other datasets however is still significant.
More importantly, in Figure~\ref{fig:percolation_cut}(b) we plot the fraction of edges cut whose weight (or IET) is less than the interval size.
This captures edges that are cut which should not be cut, assuming we believe the interval-width is the timescale over which events should be considered connected.
Interestingly \texttt{random-complete} and \texttt{academic-coauthors} display a non-monotonic relationship with the interval-width within this range.
In particular, splitting \texttt{academic-coauthors} into single year intervals results in the maximum number of edges cut (at just over 20\%).
This suggests that an intermediate interval-width is worse than a more extremal interval-width.
The other datasets have a much smaller average IET to total time period ratio and so it is anticipated that they will also display non-monotonic behaviour over a larger range of $w$.

\subsubsection{Temporal Component Clustering}

Temporal components have been the focus of recent work which captures the behaviour of individuals and collectives in temporal networks~\cite{mellor2018classifying}.
This work characterises temporal components using both features of the event graph (empirical IET and temporal motif distributions) as well as features of the induced aggregate graph (clustering, reciprocity) to capture higher-order relations.
Here we show only the outcomes of component clustering for two datasets, although implementation details can be found in~\cite{mellor2018classifying}.

\begin{figure}[!h]
	\centering
	\begin{subfigure}{0.24\linewidth}
		\centering
		\includegraphics[width=\linewidth]{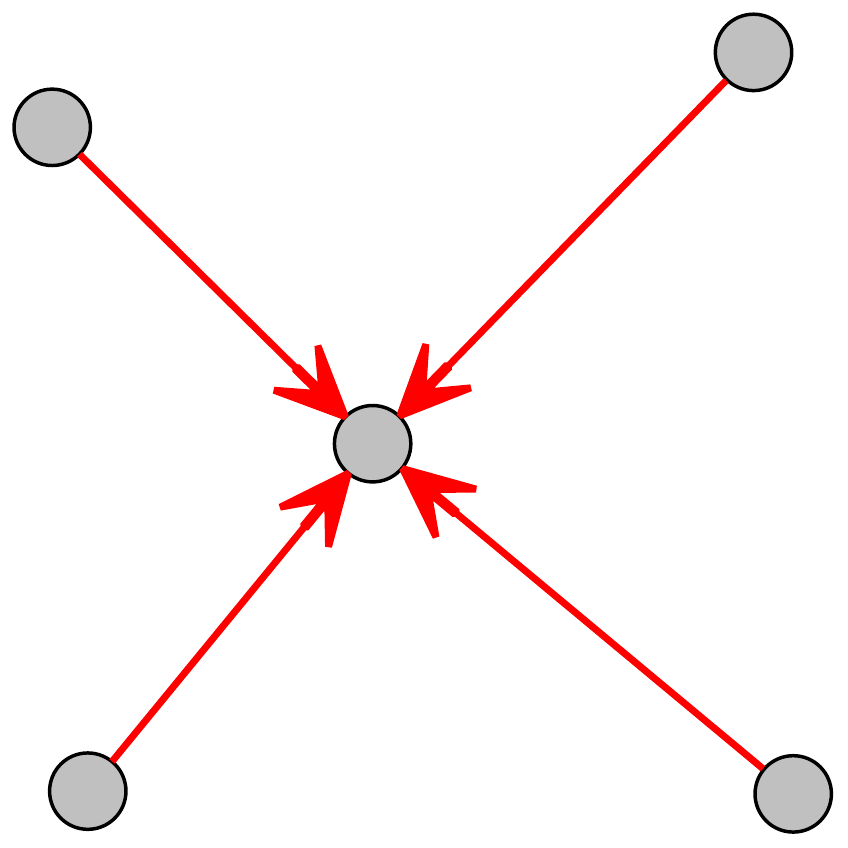}
		\caption{}
	\end{subfigure}
	\begin{subfigure}{0.24\linewidth}
		\centering
		\includegraphics[width=\linewidth]{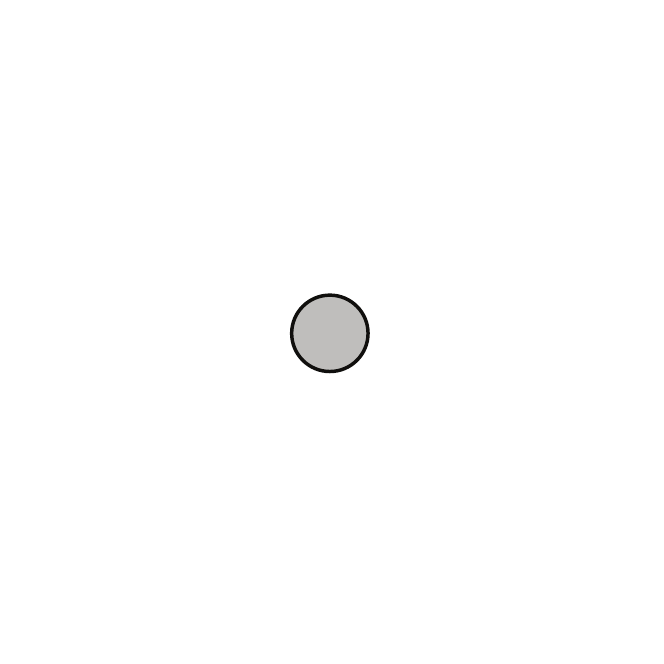}
		\caption{}
	\end{subfigure}
	\begin{subfigure}{0.24\linewidth}
		\centering
		\includegraphics[width=\linewidth]{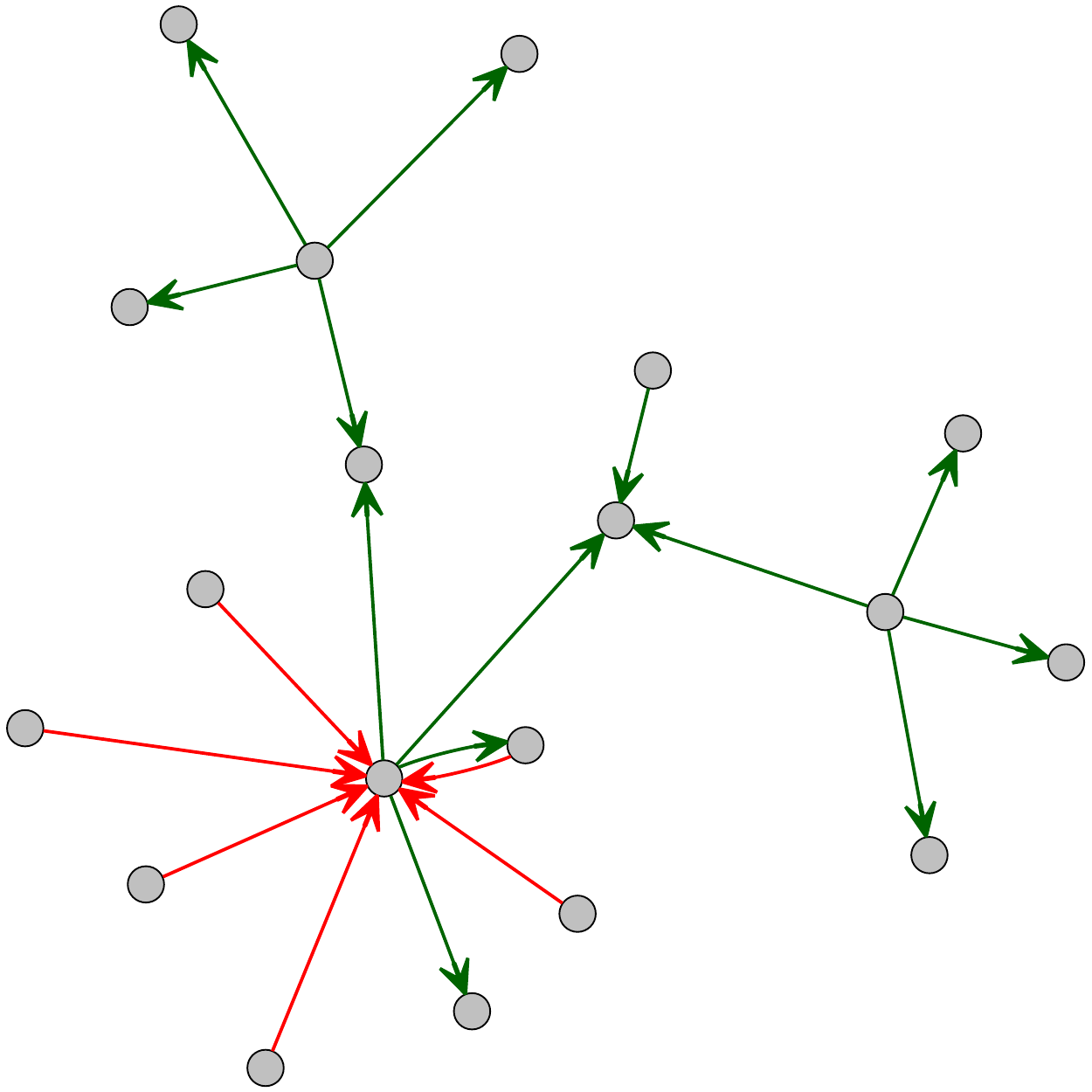}
		\caption{}
	\end{subfigure}
	\begin{subfigure}{0.24\linewidth}
		\centering
		\includegraphics[width=\linewidth]{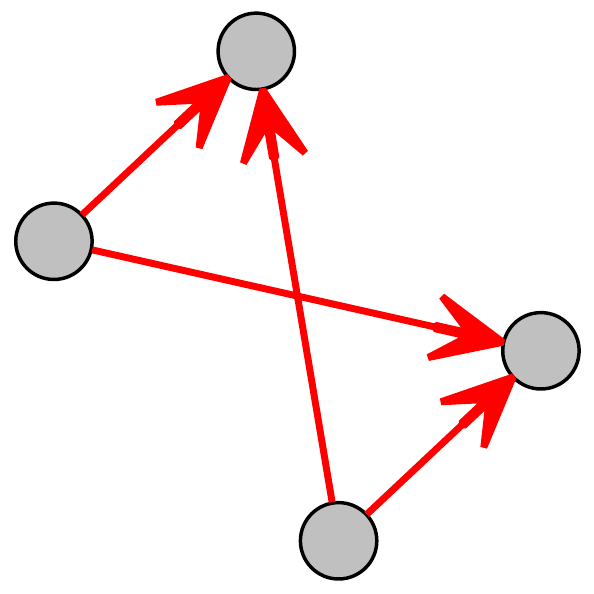}
		\caption{}
	\end{subfigure}
	\vspace{1em}

	\begin{subfigure}{0.24\linewidth}
		\centering
		\includegraphics[width=\linewidth]{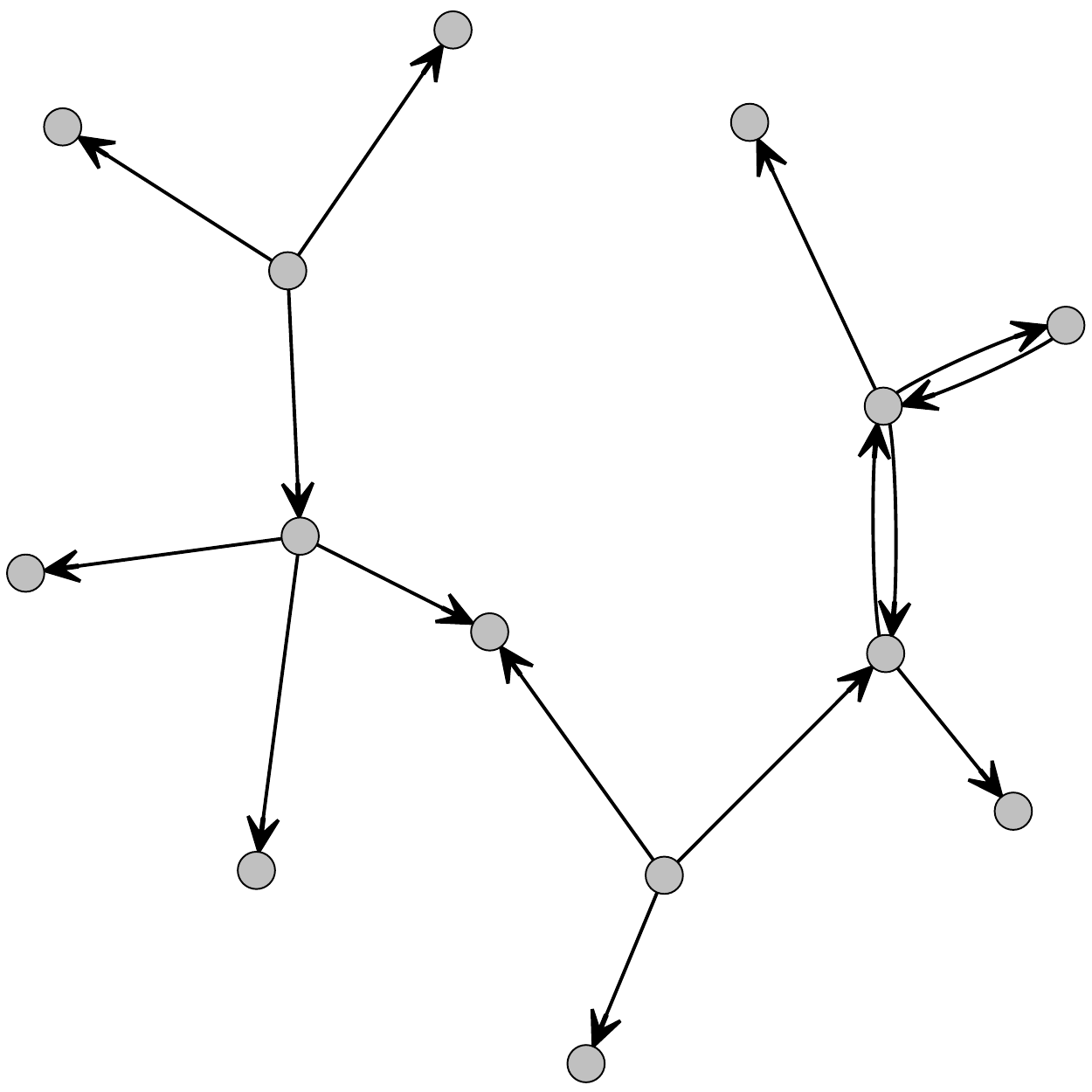}
		\caption{}
	\end{subfigure}
	\begin{subfigure}{0.24\linewidth}
		\centering
		\includegraphics[width=\linewidth]{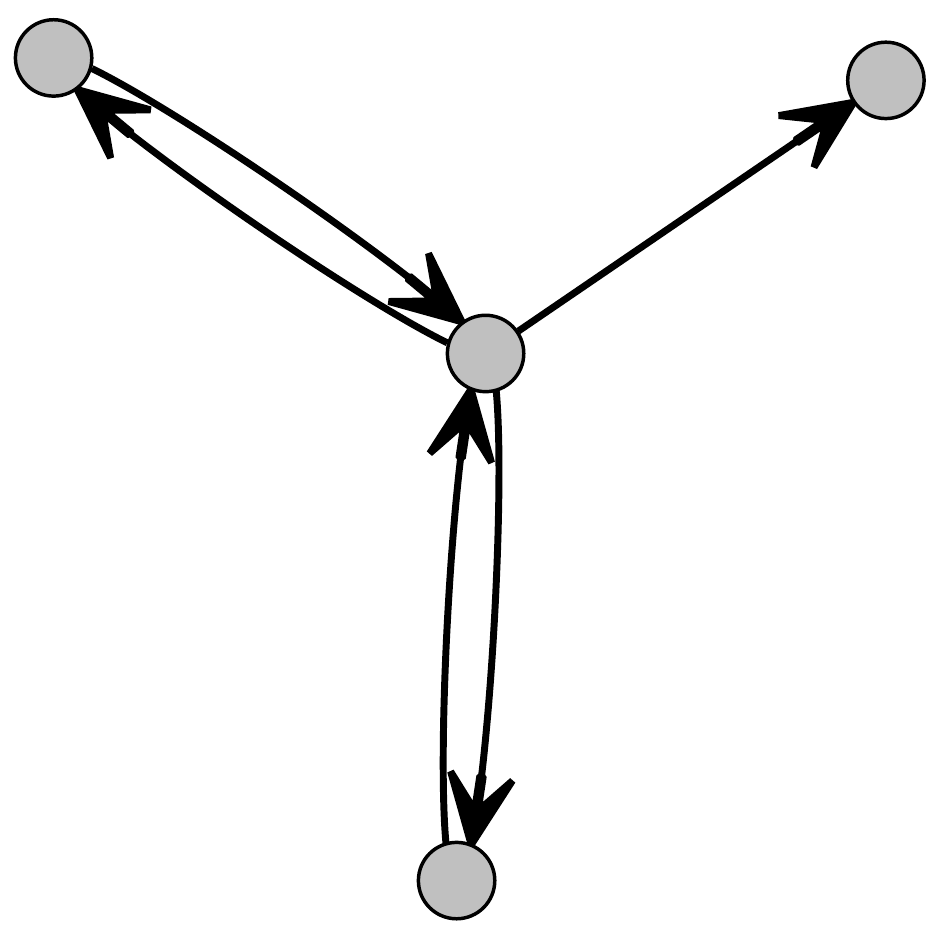}
		\caption{}
	\end{subfigure}
	\begin{subfigure}{0.24\linewidth}
		\centering
		\includegraphics[width=\linewidth]{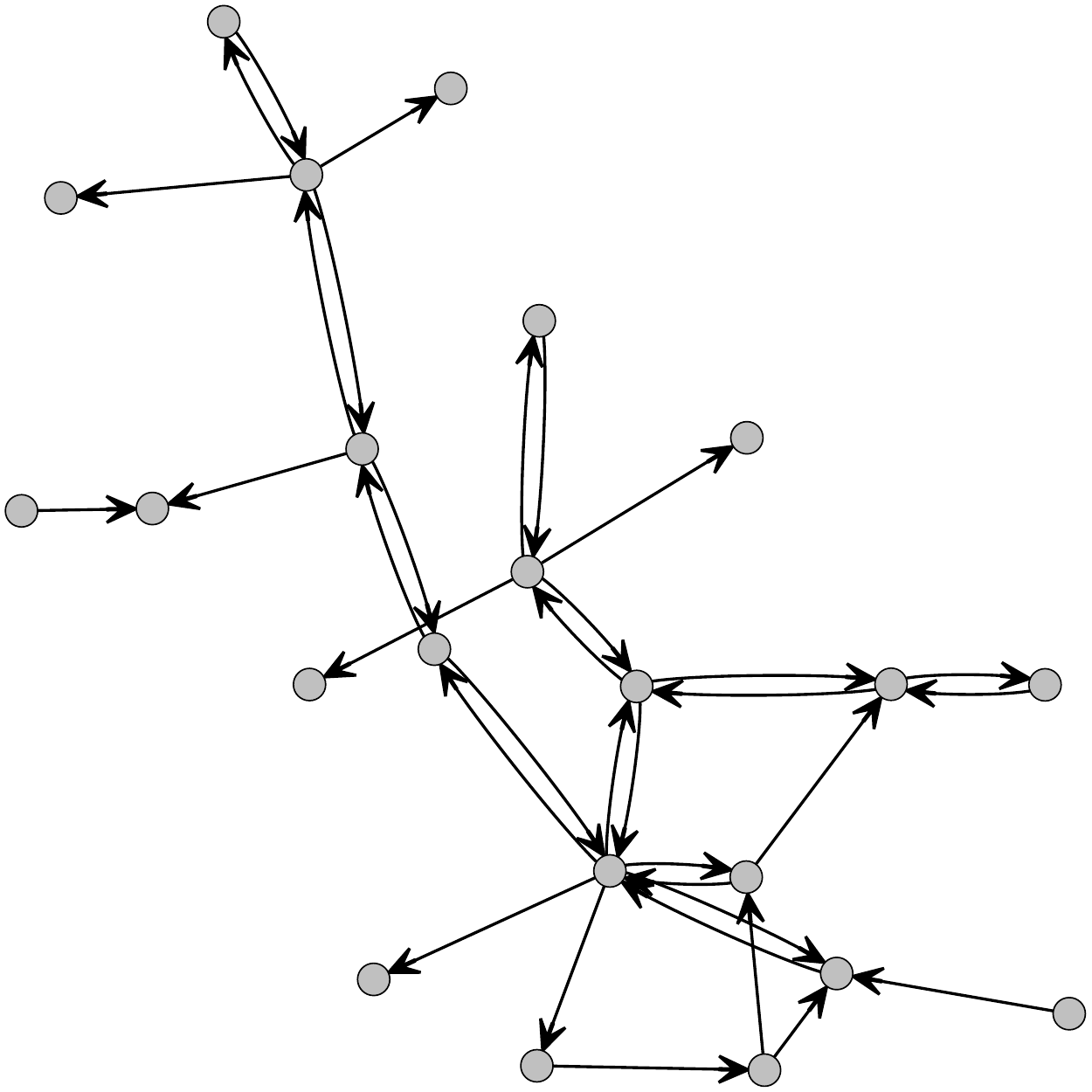}
		\caption{}
	\end{subfigure}
	\caption{
		Aggregate graph representations of representative temporal components taken from assigned clusters for the \texttt{twitter-emirates} dataset (a)-(d) and the \texttt{social-ucirvine} dataset (e)-(g).
		The \texttt{twitter-emirates} clusters correspond to (a) serial retweeting by a central account, (b) series of statuses (with no external interaction), (c) a mix of retweeting and messaging, and (d) multiple users being retweeted by multiple other users.
		Here green edges correspond to messages and red edges correspond to retweets, with the arrow representing the direction of information flow.
		In contrast, the behavioural clusters from the \texttt{social-ucirvine} dataset are (e) primarily unreciprocated broadcasting, (f) localised reciprocated messaging, and (g) reciprocated chains.
	}
	\label{fig:clustering_examples}
\end{figure}

Figure~\ref{fig:clustering_examples} shows examples of the induced aggregate graph representations of temporal components taken from unsupervised behavioural clusters. 
In (a)-(d) we have four types of collective behaviour found in the \texttt{twitter-emirates} dataset.
These behaviours span from individual-focused behaviour such as retweeting (a) and posting individual messages which receive no external response (b), to wider conversations (c) and shared retweeting (d).
These clusters are able to capture both lower-order (single posts) and higher-order interactions (multiple messaging) which would be lost by considering a network view only.
The \texttt{social-ucirvine} network (e)-(g) exhibits different behaviour to that of \texttt{twitter-emirates} as edges are much more likely to be reciprocated.
Using the same methods, this time on dyadic and unlabelled interactions we can discern three different behaviours: (e) broadcasting, (f) small reciprocated messaging, and (g) long reciprocated chains of messages.

The analysis of temporal components shows promise as an alternative to fixed-width interval temporal decomposition.
Prescribing our assumptions on the time between events means that we do not risk omitting any pair of connected events. 
However one caveat of this method is that we lose the ability to use established matrix methods as we can no longer represent the network as a sequence of static adjacency matrices.

\subsection{Event Centrality}
\label{subsec:centrality}

Much like edge centralities for static networks (e.g. edge betweenness \cite{girvan2002community}) we can define new centralities for temporal events.
Event centralities can be used to understand the importance of an event in terms of its ability reach to nodes and events further forward in time, or how well it creates a bridge between otherwise disjoint sets of nodes.
While there are many possibilities for defining centrality, here we consider an application of the communicability centrality \cite{grindrod2011communicability, grindrod2014dynamical} to temporal events.
We show that not only does event centrality provide a means to rank events based on their reach to other events, but also that it uncovers patterns in node communicability which were previously hidden.

The dynamic communicability matrix \cite{grindrod2011communicability} is defined at a time $T$ by
\begin{align*}
	Q(T) = \prod_{s=0}^{T} \left[I-\alpha A(s)\right]^{-1},
\end{align*}
where $A(s)$ is the adjacency matrix at time interval $s$, and $I$ is the identity.
This definition is written in terms of a sequence of adjacency matrices however this can be reconciled with an event-based representation by considering
a time interval such that each event belongs to its own interval.
The communicability matrix $Q$ counts all weighted temporal walks from one node to another, where the traversal of each edge is given a multiplicative weight $0<\alpha<1$.
Temporal walks can also be weighted by their age through the \emph{running} dynamic communicability matrix $S(t)$, defined iteratively by
\begin{align*}
	S(t) = \left[I + e^{-\beta \delta t} S(t-1)\right]\left[I - \alpha A(t)\right]^{-1} - I, 
\end{align*}
where $\beta$ is the downweighting factor, and $\delta t$ is the time between intervals. 

Now consider a directed temporal network of $m$ events, $n$ nodes, and the corresponding event graph with a walk-forming joining rule (Section~\ref{sec:event_graph}).
That is, each event in the graph is connected to all future events with which it forms a two-step walk.
Let $A_{\rm EG}$ be the adjancency matrix representation of such an event graph, i.e,
\begin{align*}
	\left(A_{\rm EG}\right)_{ij} = \begin{cases}
	1 & \text{ if } f(e_i, e_j)\\
	0 & \text{ otherwise.}
\end{cases}
\end{align*} 
The event communicability matrix is then given by $M = \alpha(I-\alpha A_{\rm EG})^{-1}$ where we have included an extra prefactor of $\alpha$ due to network being second order.
Furthermore we can incorporate the temporal decay of walks by considering a weighted adjacency matrix $A^*_{\rm EG}$
which takes the form
\begin{align*}
   \left(A^*_{\rm EG}\right)_{ij} = \begin{cases}
        e^{-\beta (t_j - t_i)} & \text{ if } f(e_i, e_j) \\
        0 & \text{ otherwise.}
    \end{cases}
\end{align*}
The event communicabiity matrix $M^*$ is constructed as before, replacing $A_{\rm EG}$ with  $A^*_{\rm EG}$.

The sum of weighted walks from each event $i$ (sometimes referred to as the broadcast centrality) is given by $\vec{b} = M \vec{1}_m$ where $\vec{1}_s = (1,\dots,1)$ is a vector of dimension $s$.
For temporal decay we can consider two centralities, either by the right multiplication of $M^*$ by $\vec{1}_m$ or by $d(T)$ where  
\begin{align*}
    d_i(T) = e^{-\beta(T-e_i)}.
\end{align*}
The former downweighs walks in time only for the duration of the walk, while the latter downweighs walks in time even after they have terminated (which aligns with the running dynamic communicability).
We may interpreted this difference as ``how much information has been transferred from event to event, assuming loss over time and at each transmission?'', in contrast to the running dynamic communicability which is ``how much information has been transferred from event to event recently?''
Other centralities can be created by considering the node-event-source and node-event-target incidence matrices $X_{s}, X_{t}$, given by
\begin{align*}
    (X_{s})_{ij} = \begin{cases}
        1 & \text{ if } i=u_j \\
        0 & \text{ otherwise},
    \end{cases} \\
    (X_{t})_{ij} = \begin{cases}
        1 & \text{ if } i=v_j \\
        0 & \text{ otherwise}
    \end{cases} 
\end{align*}
respectively, which state whether a node is either a source or a target in an event.
For example, the number of walks from an event $i$ to node $j$ is given by the $(i,j)$ entry of $M X^T_t$.
This can be used to calculate how many nodes an event may potentially influence in the future.
Finally, we can recover the dynamic communicability matrix as $Q = X_s M X^T_t$ (a sketch proof is given in Appendix~\ref{app:communicability}).
Writing the broadcast centrality for nodes as $\vec{b} = X_s M \vec{1}_n$ shows that a node is only as central as the events it participates in.
There is therefore extra information in this representation as we can consider the distribution of event centralities for which a node participates as well as its overall centrality.

The need to consider event centrality can be seen with a simple example.
Consider a random temporal network of $n$ nodes that is generated in discrete time.
At each iteration $t$, a source node $u$ is picked uniformly at random from $\{1,\dots,n\}$, and and a target $v$ picked uniformly from $\{1,\dots,n\} \setminus {u}$ to create a temporal event $(u,v,t)$.
We consider a special node $u^*$, such that if an event $(u^*,v,t)$ occurs for some $v,t$, the subsequent event will be $(v,w,t+1)$ for some $w$.
This way we create a node with a walk-forming advantage as there is guaranteed to a walk of length two for every event that node $u^*$ initiates, and this walk will occur quickly (at the next time step).
We anticipate that we should be able to easily identify this node from the broadcast centrality.

We consider the above model, run for $n=20$ nodes and $m=1000$ events.
Using the time-decay centrality (but not the running dynamic centrality)\footnote{
    This way we eliminate the issue of where we `stop' our analysis as in the running dynamic centrality $b(t)$ is a continuous function of time.
} we calculate the broadcast scores of all nodes in the network.

\begin{figure}[!h]
	\centering
	\includegraphics[width=0.4\linewidth]{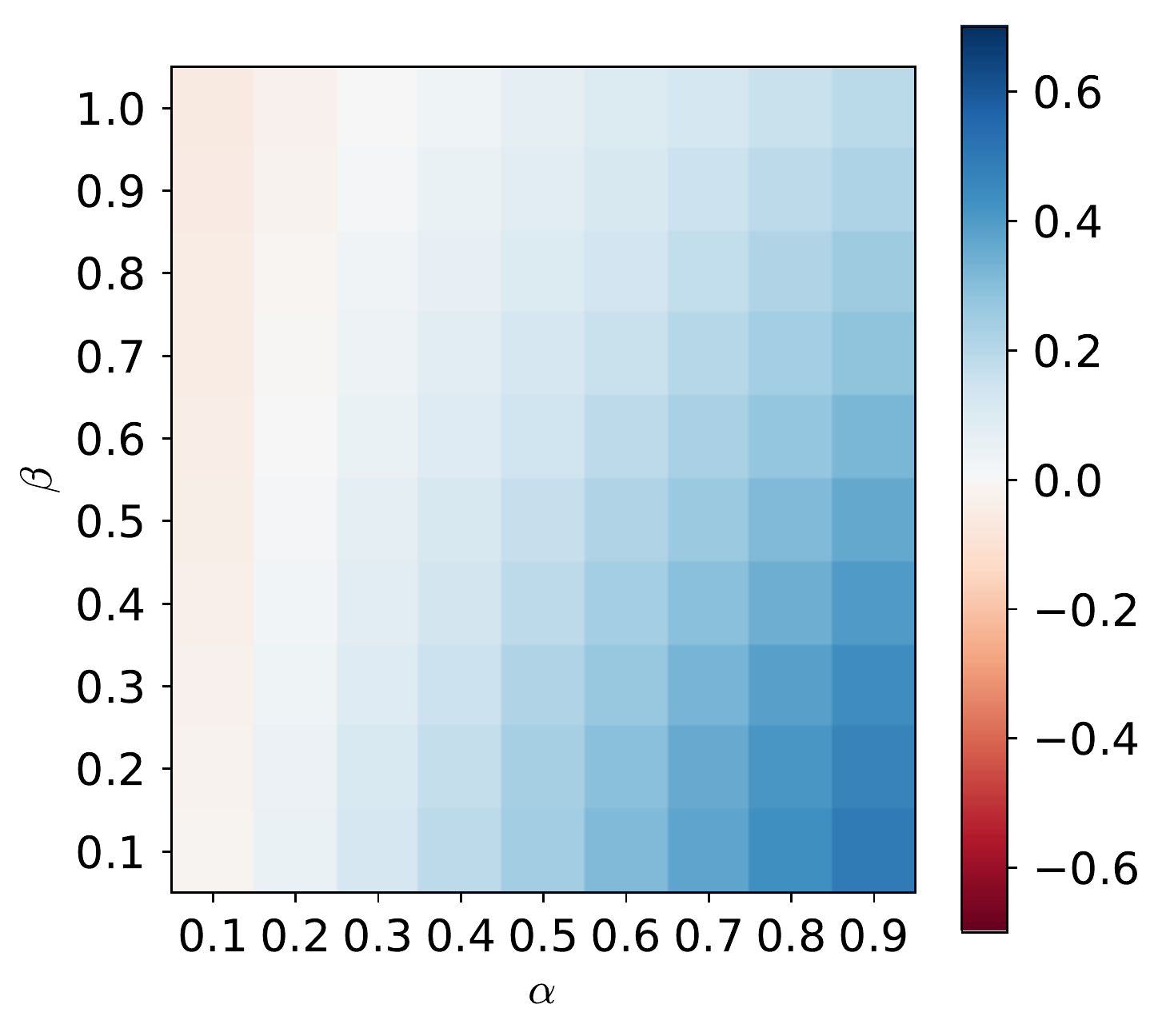}
	\caption{
		Percentage difference in broadcast score between the node $u^*$ and the competitor node $u^c$.
		For large values of $\alpha$, i.e. $\alpha>0.5$, the node $u^*$ has a significant advantage over the nearest competition as one would expect.
		However for smaller values of $\alpha$ (and in particular for high values of $\beta$), this advantage is drastically reduced, and the two nodes are either indistinguishable or the competitor prevails.}
	\label{fig:communicability_parameters}
\end{figure}

In Figure~\ref{fig:communicability_parameters} we plot the percentage difference in broadcast centrality between node $u^*$ and the next nearest competitor node $u^c$.
To elaborate, if node $u^*$ is the most central, then we compare with the second most central node.
If node $u^*$ is not the most central node, we compare it with the most central node.
Here, we see that there is a range of $\alpha$ and $\beta$ for which node $u^*$ is the most prominent node (blue), but there are regions of parameter space where either it is indistinguishable from another node (white), or it is not the most central node (red).
This is potentially problematic for non-synthetic datasets where we do not know exactly how the choice of $\alpha$ and $\beta$ affects the centrality ranking. 

\begin{figure}[!h]
	\centering
	\includegraphics[width=0.7\linewidth]{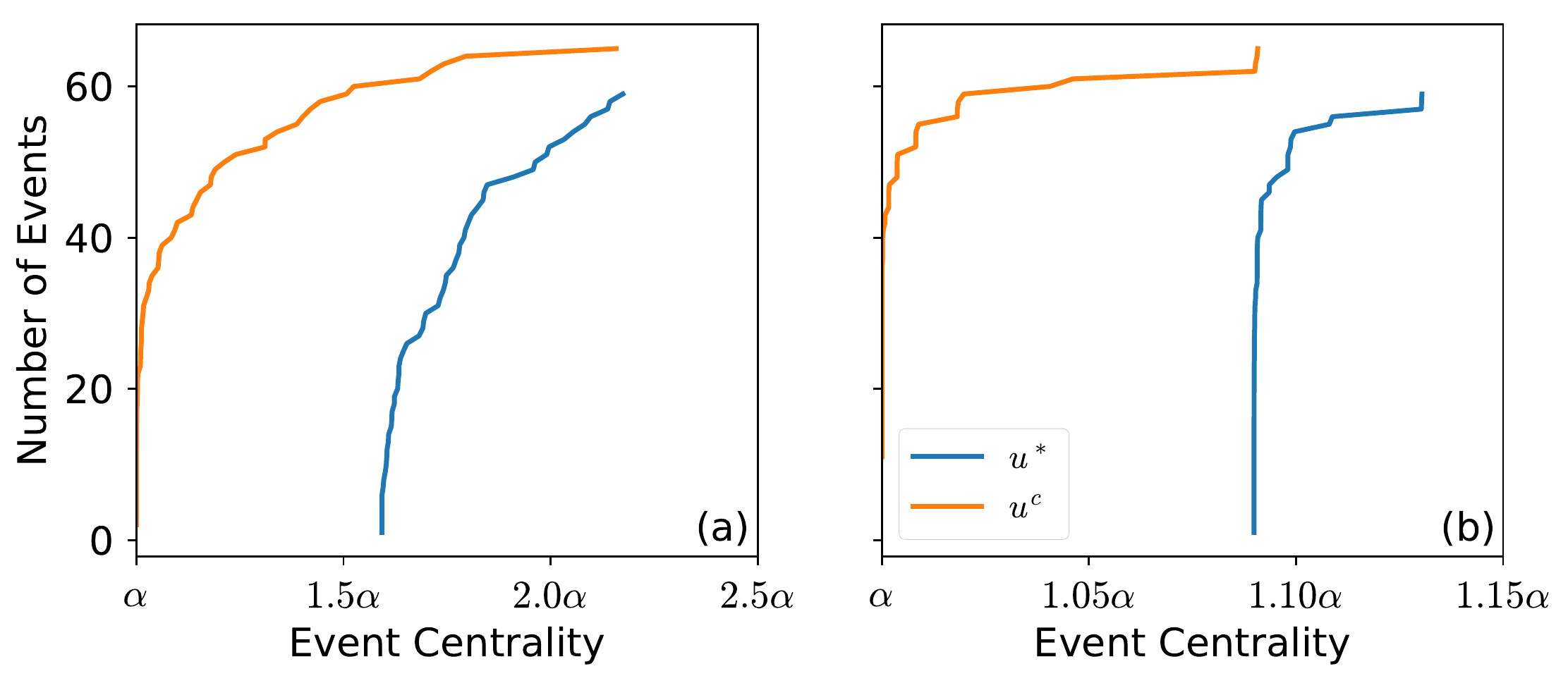}
	\caption{
		The cumulative distribution of event centralities for the the special node $u^*$ and the challenger node $u^c$ for two sets of parameters.
		In (a), where $(\alpha,\beta)=(0.8,0.3)$, we see that $u^*$ has participated in fewer events than the closest challenger however the events that $u^*$ participates in have at least a centrality of $\alpha + \alpha^2$.
		Despite participating in more events than $u^*$, the majority of events for node $u^c$ have a centrality less than $\alpha + \alpha^2$, resulting in a higher broadcast score for node $u^*$.
		In contrast for (b), where $(\alpha,\beta)=(0.2,0.8)$, despite all events for $u^*$ being more central than any of those for $u^c$, the difference between the event centralities is outweighed by the number of events, and therefore node $u^c$ is deemed to be the most central.
	}
	\label{fig:communicability_event_distributions}
\end{figure}

To remedy this situation we can examine the distribution of event centralities that each node participates in.
In Figure~\ref{fig:communicability_event_distributions} we examine the distribution of event centralities for node $u^*$ and the closest competitor for two different parameter regimes; one where node $u^*$ is the most central (a), and one where it is not (b).
In both cases the competitor node participates in more events than $u^*$.
However in both cases, we can identify node $u^*$ as the minimum event centrality is greater than the majority of event centralities for the other node. 
This means that node $u^*$ is participating in fewer but important events than $u^c$.
So how does $u^*$ lose the top ranking? In Figure~\ref{fig:communicability_event_distributions}(b) node $u^*$ is still participating in more important events, however \emph{the relative extent of importance} of those events has diminished.
Comparing the scale of the x-axis in both subplots, in (a) the most important event is over twice as important as the least. 
In (b) however this is reduced drastically and the most important event is less than $15\%$ more important than the least.
In this scenario, when all events are approximately equal, the number of events a node participates in becomes most important.

\subsection{Higher-order Graphical Models}

Event graphs can be used to analyse and describe empirical temporal networks, but they do not provide a model of the temporal network which we can use to sample new networks.
Ideally we would like to make predictions about subsequent events or to generate new synthetic networks which are structurally and temporally similar.
As we saw in Section~\ref{sec:introduction}, the multi-order model of Scholtes~\cite{scholtes2017network} is capable of exactly this.
Focusing on the second-order model, given that an event is observed, the probabilistic model gives the likelihood of each event occurring afterwards.
Furthermore to generate a new temporal network from the model one can record the journeys of one or more random walkers on the network.
To explicitly generate this second-order model from data, we first construct the event-subsequent $\Delta t$-adjacent event graph.
We can then construct the graphical model by calculating for each node $(u,v)$ the proportion of times an event $(w,z)$ follows it, i.e. the number of times an edge appears in the event graph between events $(u,v,t_1)$ and $(w,z,t_2)$ for any $t_1,t_2$.
This in effect \emph{flattens} the time dimension of the event graph.

The relationship between the event graph and these models is clear, but it also motivates other more complex models.
The higher-order model in~\cite{scholtes2017network} described previously can predict the sequence of events, but not model the times between them occurring.
Since the IETs are captured as edge weights in the event graph we can use them to further inform the flattened model\footnote{
	We cannot fully infer an inter-event time distribution from a finite size time window however, see~\cite{kivela2015estimating} for more details.
}.
Now, as well as a probability of observing a subsequent event we also know the distribution of times between those events from which we can sample.
Alternatively we can use only the IET distributions and move our random walker to the event that occurs first.
This is a similar approach to that of the `waiting time' of walkers in~\cite{lambiotte2013burstiness}.

\section{Conclusion}
\label{sec:conclusion}

In this article we have introduced and generalised the notion of an event graph, a second-order time-unfolded graphical model of a temporal network.
We have also extended this definition to include non-dyadic interactions (known as hyper-events), and shown that using the event graph representation both dyadic and non-dyad interactions can be be modelled in the same way.
We have highlighted a number of existing works of where event graphs have been used to uncover the structure of temporal networks and have offered more speculative examples of domains such as motif counting, centrality, and network decomposition where this type of second-order model shows utility.

Event graphs are not without their limitations.
There are considerations on the computational complexity of the event graph construction.
As a second-order model there are as many nodes in the event graph as there are events.
Typically there are significantly more events than nodes in the original network (unless each node interacts only once).
The increase in the size of the graph considered can be counteracted by employing efficient algorithms that exploit the fact that the event graph is a directed acyclic graph.
The computational complexity of generating an event graph is heavily dependent on the joining function.
For node-subsequent adjacency an event can be added to the event graph in an $\mathcal{O}(1)$ operation which makes this rule suitable for real-time analysis of event streams.
Conversely, a simple adjacency rule requires $\mathcal{O}(m)$ operations for the addition of single event where $m$ is the number of previous events.
For small networks this is may be feasible, however for larger networks a restriction on $\Delta t$ may be needed.

There is also further difficulty choosing the parameter $\Delta t$, however this choice will become easier as we understand more about the timescales involved in temporal networks. 
Furthermore there are unaddressed issues with temporal networks which have multiple timescales and how these should be modelled in generality.
We first need to develop methods that can detect these timescales and then implement new joining rules to accommodate them.

As the data we are collecting is becoming more complex, and in particular more \emph{relational}, we are seeing a shift away from simplistic first-order models towards second- and higher-order models.
In this article we make some progress in both senses of the term `higher-order', modelling both second-order dependences and non-dyadic interactions.
There are however still many more avenues to explore to aid our understanding of temporal networks and to create meaningful and computationally feasible higher-order models.

\subsubsection*{Code Availability}

Code available freely online as part of the \texttt{eventgraphs}\footnote{
	\url{https://github.com/empiricalstateofmind/eventgraphs}
} Python package~\cite{url_eventgraphs}.
Notebooks which cover the analysis of this article are available in the \texttt{examples/advances\_and\_applications\_paper} folder of the repository.


\appendix

\section{Event and Node Communicability Centrality}
\label{app:communicability}

\textbf{Claim:} 
We claim that the event and node dynamic communicability matrices ($M$ and $Q$ respectively) are related by $Q = X_s M X^T_t$, where $X_s$ and $X_t$ are the node-source and node-target incidence matrices respectively (defined in the main text).

\textbf{Proof (sketch):}
To prove this we show that the matrix $M$ captures all temporally weighted walks in the temporal network.
Suppose there is a temporal path of length $k$ from node $u^*$ to node $v^*$.
Then by definition, there exists a sequence of timestamped edges (or events)
\begin{align*}
    (n_1, n_2, t_1), (n_2, n_3, t_2), \dots, (n_k, n_{k+1}, t_k)
\end{align*}
such that $n_1 = u^*$ and $n_{k+1} = v^*$ and $t_i < t_{i+1}$ for all $i$.
For ease, label these events $e_1, e_2, \dots, e_k$.

We now construct the event graph matrix $A_{\rm EG}$ of these events. 
For all $i$, $(A_{\rm EG})_{i,i+1} = 1$ as $v_i = n_{i+1} = u_{i+1}$ and $t_{i+1} - t_i > 0$.
The graph consists of a single chain of length $k-1$, $e_1 \to e_2, \to \dots \to e_k$ assuming that all nodes $(n_i)_{i=1}^{k+1}$ are distinct\footnote{
    If they are not distinct then there potentially will be other edges and other possible walks, although none will have length greater than $k-1$.
}.
This implies $(A_{\rm EG}^{k-1})_{1k} = 1$ with $u_1=u^*$ and $v_k=v^*$.
Since the chosen walk is arbitrary $\alpha^k A_{\rm EG}^{k-1}$ will capture all weighted walks of length $k$.

Therefore, to capture all walks we consider
\begin{align*}
    M:= \alpha(I - \alpha A_{\rm EG})^{-1} = \alpha I + \alpha^2 A_{\rm EG} + \alpha^3 A_{\rm EG}^2 + \dots
\end{align*}
noting that $A_{\rm EG}^0=I$.
We can guarantee convergence as the event graph is a DAG, so there exists an integer $R$ such that $A_{\rm EG}^R = 0$. 
We can therefore chose $\alpha$ freely in $[0, \infty)$ (although we choose $\alpha \in (0,1)$ typically).
The matrix $M$ acts similarly to the matrix $Q$ in that it contains the information of all walks in the network.
To find all walks which start at a particular node, we premultiply by $X_s$. 
The non-zero entries in each row of $X_s M$ give the contribution of each event to the node.
By a similar argument we can show that $M X_t^T$ counts the weighted number of walks from each event to each node.
Combining these we can calculate the weight of walks that start at one node and end at another, recovering the node communicability matrix as
\begin{align*}
    Q = X_s M X_t^T.
\end{align*}

\end{document}